\documentclass[a4paper,12pt,onecolumn,notitlepage]{article}
\usepackage[pdfpagemode=None,bookmarksopen=false,colorlinks=true,linkcolor=blue,citecolor=blue]{hyperref}
\usepackage{amsmath}
\usepackage{amssymb}
\usepackage{sectsty}
\usepackage{bibspacing}
\usepackage{rpkid}
\allsectionsfont{\normalsize}
\begin{document}

%-------------------------------------------------------------
\thispagestyle{empty}
\begin{center}
\noindent
{\Large \textbf{Quantum theory as inductive inference}}\footnote{\scriptsize Published in: Mohammad-Djafari A., Bercher J., Bessi\`{e}re P. (eds.), \textit{Proceedings of the 30th International Workshop on Bayesian Inference and Maximum Entropy Methods in Science and Engineering}, AIP Conf. Proc. \textbf{1305} (2011), Springer, Berlin, pp. 24-35. The published version was abridged due to strict 8 pages article size limit. This text is an improved version of an original unabridged paper.}\\% Full version is available at: \href{http://www.arxiv.org/pdf/1009.2423}{arXiv:1009.2423}.}\\
\ \\
{Ryszard Pawe{\l} Kostecki}\\
{\small\ \\}{\small
\textit{Institute of Theoretical Physics, University of Warsaw, Ho\.{z}a 69, 00-681 Warszawa, Poland}}{\small\\
\ \\
September 11, 2011}%August 30, 2011}%August 21, 2011}
\end{center}
%-------------------------------------------------------------
\begin{abstract}{\small\noindent We present the elements of a new approach to the foundations of quantum theory and probability theory which is based on the algebraic approach to integration, information geometry, and maximum relative entropy methods. It enables us to deal with conceptual and mathematical problems of quantum theory without any appeal to frameworks of Hilbert spaces and measure spaces.} 
%{\scriptsize PACS: 89.70.Cf 02.50.Cw 03.67.-a 03.65.-w}
\end{abstract}
\section{Introduction}
%------------------------------------------------------------
What is the relationship between information theory and quantum theory? The notion of information, quantified by entropy, is usually defined using the notion of probability \cite{Shannon:1948}. But Ingarden and Urbanik \cite{Ingarden:Urbanik:1961,Ingarden:Urbanik:1962} showed that the notion of information is independent of the notion of probability, and that the latter can be defined by the former. Moreover, the mathematical setting of probability theory can be considered as a special case of the mathematical setting of quantum theory, with Hilbert spaces \cite{Gleason:1957,Piron:1976,Randall:Foulis:1983,Wilce:2000,Pitowsky:2006} or $W^*$-algebras \cite{Maeda:1990,Hamhalter:2003,Redei:Summers:2007} replacing the role played by the measure spaces $(\X,\mho(\X),\mu)$ in the Borel--Kolmogorov approach. Hence, in order to consider information theory as more fundamental than quantum theory, it is necessary to develop it independently of the notion of probability, and to equip it with the dynamical structures sufficient to recover dynamics of quantum theoretic models. 

In this paper we will present the elements of a new mathematical formulation of information kinematics and information dynamics, equipped with a new interpretation, which are aimed to solve the above task. We begin with replacing the ordinary Borel--Kolmorogov setting based on probability measures $\mu$ on commutative countably additive algebras $\mho(\X)$ of subsets of a given set $\X$ by \textit{information kinematics} based on spaces of finite positive integrals on abstract (but \textit{integrable}) commutative or non-commutative algebras. This change has two reasons:
\begin{enumerate}
\item[(1)] the conflict between the Bayes--Laplace \cite{Bayes:1763,Laplace:1812} and the Borel--Kolmogorov \cite{Borel:1909:denom,Kolmogorov:1933} approaches to mathematical foundations of the probability theory can be resolved in terms of the Le~Cam--Whittle \cite{LeCam:1964,LeCam:1986,Whittle:1970,Whittle:2000} approach, which is based on the Daniell--Stone \cite{Daniell:1918,Daniell:1919:a,Daniell:1920,Stone:1948,Stone:1949} theory of integration on vector lattices, using the canonical association of a vector lattice to a given boolean algebra;
\item[(2)] the algebraic approach to mathematical foundations of quantum theory \cite{Segal:1947:postulates,Segal:1953,Haag:1992}, provided in terms of positive linear functionals on non-commutative $C^*$-algebras avoids several important problems of the Hilbert space based approach, including the problem of unitary inequivalence of different Hilbert space representations of a single non-commutative algebra in infinite-dimensional cases, which plays a crucial role in construction of mathematically strict models in relativistic quantum field theory and continuous quantum statistical mechanics \cite{Emch:1972,Bratelli:Robinson:1979,Haag:1992,Baumgaertel:1995,Araki:1999}.
\end{enumerate}

In order to provide a refined analysis and characterisation of the information kinematic spaces of integrals, we equip them with the non-symmetric information distance function, known as  \textit{information deviation} (negative \textit{relative entropy}). Relative entropy defines the \textit{information geometry} of information kinematic spaces in terms of such objects as riemannian metrics, affine connections, and non-linear projections (on the convex subsets), which can be naturally derived from relative entropy under some mild conditions. We discuss the characterisation of a particular family ($D_p$) of relative entropy functionals. As a consequence of this characterisation, some preferred class of information geometries is selected. We introduce the \textit{information dynamics} as a mapping on the space of finite integrals provided by the constrained maximisation of relative entropy. 

When applied to the abstract commutative boolean algebras, the above setting generalises the standard approach to probability theory \textit{and} statistical inference. When applied to the abstract non-commutative $W^*$-algebras, the above setting provides new mathematical framework for kinematics and dynamics of quantum theory. In both (commutative and non-commutative) regimes we are able to recover the standard approaches to foundations as special cases of our approach. 

The new kinematical setting of quantum theory is defined as follows. Given $W^*$-algebra $\N$, the \textit{quantum information model} $\M(\N)$ is defined as (some) subset of the space $\N_*^+$ of normal positive finite linear functionals on $\N$. The space $\M(\N)$, together with its non-linear quantum information geometry, is considered as a replacement of the linear Hilbert space $\H$ in the role of the kinematic setting of a quantum theoretical model. The elements of $\M(\N)$ will be called \textit{quantum information states}. The family $D_p$ of quantum relative entropies is closely related with non-commutative $L_p(\N)$ spaces (because $D_p$ are quantum Bregman entropies), so the preferred geometry of $\M(\N)$ induced by $D_p$ can be analysed by quantum information geometric representations of abstract $\M(\N)$ spaces in non-commutative $L_p(\N)$ spaces. This replaces the use of commutative $L_2(\X,\mho(\X),\mu)$ spaces which represent the abstract Hilbert spaces $\H$. Among $L_p(\N)$ spaces, of particular importance is the $L_2(\N)$ space, which is unitary isomorphic to the Hilbert space $\H_H$ of Haagerup's standard representation of $\N$, and unitary isomorphic to the Hilbert space $\H_\omega$ of the Gel'fand--Na\u{\i}mark--Segal representation associated with any faithful $\omega\in\N_*^+$. The quantum information geometric representations $\ell_{1/2}:\M(\N)\ra L_2(\N)$ serve as a key feature allowing the reconstruction of the ordinary Hilbert space based setting of quantum theory as a special case of our approach.

The new dynamical setting of quantum theory is defined as follows. The description of temporal behaviour and---more generally---information dynamics of quantum models in terms of unitary or completely positive mappings of operators over $\H$ is replaced by the mapping on $\M(\N)$ generated by the constrained maximisation of quantum relative entropy $D_p$. This map can be understood in terms of information geometry as a non-linear projection. Its well-defined constraints (guaranteeing the existence and uniqueness of the solution of the variational problem) are specified by non-empty closed convex subsets of non-commutative $L_p(\N)$ spaces. These subsets define the domain of the entropic projection. If the temporal dependence of the constraints is reflected in the temporal dependence of the projection, then this projection generates a temporal trajectory on $\M(\N)$, representing the \textit{temporal evolution} of quantum information states.

Definining quantum kinematics in terms of spaces $\M(\N)$ of integrals over abstract (qualitative) non-commutative $W^*$-algebras $\N$ instead of dealing with Hilbert space representation-dependent density operators and von Neumann algebra representations of abstract $W^*$-algebras provides novel perspective on the foundations of quantum theory. This approach directly bypasses the problems of representation-dependent separation of the Hilbert space vectors into amplitudes and phases that has troubled the attempts to consider the orthodox Hilbert space approach to quantum theory as an extension of probability theory. More generally, the problem of contextuality of probabilistic predictions derived from orthodox approach that are implied by the restriction of spectral representation to commutative subalgebras plays no role in our approach. 

On the mathematical level, our approach follows the ideas of an algebraic approach to quantum theory developed by Segal, Haag and others \cite{Segal:1947:postulates,Segal:1953,Haag:1959} (for reviews see \cite{Emch:1972,Bratelli:Robinson:1979,Haag:1992,Baumgaertel:1995,Araki:1999}). On the conceptual level, we follow the ideas of Jaynes \cite{Jaynes:1957,Jaynes:1957:2,Jaynes:1967,Jaynes:1979:where:do:we:stand,Jaynes:1986,Jaynes:1990,Jaynes:2003} on the character of statistical mechanics and quantum theory as probabilistic statistical inference theory, and the ideas of Ingarden \cite{Ingarden:1963,Ingarden:1976,Ingarden:1978,Ingarden:Janyszek:1982,IJKK:1982,IKO:1997} on the foundational role of the information theory and information geometry in statistical mechanics and quantum theory. In this sense, we aim at unification of the information-theoretic and algebraic perspectives on the foundations of statistical mechanics and quantum theory.

However, as opposed to existing `objective bayesian' (e.g., \cite{Youssef:1991,Youssef:1994,Caticha:1998:PLA,Caticha:1998:PRA,Caticha:2000:insufficient}) and `subjective bayesian' (e.g., \cite{Emch:1998,Caves:Fuchs:Schack:2001,Fuchs:2003,Fuchs:2010}) derivations and/or interpretations of elements of the formalism of quantum theory, our derivation is completely independent of the notions and semantics of probability theory and Hilbert space based quantum theory. This is achieved due to the kinematic properties of our approach, which arise as an extension of an algebraic approach to quantum theory (the latter treats non-commutative algebras as fundamental entities of quantum theory). 

On the other hand, as opposed to the standard settings of quantum information theory and algebraic quantum theory, our approach can be directly used to construct dynamical (`interacting', predictive, experimentally verifiable) models of quantum field theory. This is achieved due to the dynamic properties of our approach which arise as an extension of the approaches of Jaynes \cite{Jaynes:Scalapino:1963,Mitchell:1967,Jaynes:1985:scattering,Jaynes:1985:macropred,Jaynes:1986,Jaynes:1993} and Zubarev \cite{Zubarev:1961,Zubarev:Kalashnikov:1969,Zubarev:Kalashnikov:1970:TMF,Zubarev:Kalashnikov:1970:Physica,Kalashnikov:Zubarev:1972,Zubarev:1994,Morozov:Roepke:1998} to non-equilibrium finite-dimensional quantum statistical mechanics (these approaches derive the time-dependent density operators from the constrained relative entropy maximisation with time-dependent constraints, see \cite{Grandy:1987,Zubarev:Morozov:Roepke:1996,Grandy:2008} for review).

This paper is only a brief introduction to the more detailed mathematical and conceptual discussion provided in the forthcoming series of papers \cite{Kostecki:2011:towards:series}. Its aim is to present the basic principles of the new setting in the context of mathematical formalisms of the standard setting for probability theory and quantum theory.
%-------------------------------------------------------------
\section{Towards new foundations of probability and statistical inference theory}
%-------------------------------------------------------------
\subsection{Beyond standard approaches}
%-------------------------------------------------------------
The Bayes--Laplace and the Borel--Kolmogorov approaches to foundations of probability theory can be viewed as extreme cases of an application of two competing principles: evaluational (kinematical) and relational (dynamical). The notion of probability in the Borel--Kolmorogov approach is specified by the \textit{probability measure} $\mu:\mho(\X)\ra[0,1]$ on the countably additive boolean algebra $\mho(\X)$ of (some) subsets of a given set $\X$. This notion is essentially evaluational, and it is justified by an appeal to evaluational framework provided by the measure theory. Thus, it can deal with infinite sets. However, it lacks any \textit{generic} notion of conditionalisation. On the other hand, the notion of probability in the Bayes--Laplace approach is specified by the \textit{conditional probability map} $\B\ni A\mapsto p(A|I)\in[0,1]$, which assigns a quantitative value to an element of a boolean algebra $\B$ under condition that the element $I\in\B$ has boolean value `true'. This notion is essentially relational, and it is usually justified by an appeal to framework of probabilistic statistical inference provided by Bayes' rule\footnote{Note that the Ramsey--de~Finetti type \cite{Ramsey:1931,deFinetti:1931,deFinetti:1970} or Cox's type \cite{Cox:1946,Cox:1961} derivations of the Bayes theorem (or, equivalently, of the algebraic rules of `probability calculus') \textit{assume} that the conditional probabilities $p(A|I)$ are to be used in order to draw inferences on the base of conditioned premises (evidence). Hence, they \textit{assume} that \textit{some} rule of probability updating has to be used, because only under this assumption it is possible to speak of conditioned elements of the algebra as `evidence', or to speak of conditional probabilities as `inferences'. In consequence, the use of the notion of conditional probability amounts to use of \textit{some} probability updating principle in the first place. It amounts to use of some particular algebraic rules of transformation of conditional probabilities only under \textit{additional} assumptions, which might not be relevant in the general case.}
\begin{equation}
        p(x|\theta)\mapsto p_{\mathrm{new}}(x|\theta):=p(x|\theta)\frac{p(b|x\land\theta)}{p(b|\theta)},
\label{Bayes.rule}
\end{equation}

Thus, the Borel--Kolmorogov setting is of evaluational (kinematic) character and works well with countably additive algebras, while the Bayes--Laplace setting is of relational (dynamical) character and works well with finitely additive algebras. These two foundational approaches are independent of each other and in this sense are not in any conflict. However, they generate an important \textit{decision problem}, because the Borel--Kolmogorov approach lacks any \textit{generic} notion of conditionalisation of probabilistic infrences under some evidence, while the Bayes--Laplace approach lacks any \textit{generic} extension to countably additive algebras. Thus, each of them is in some sense insufficient. We can resolve this problem by an appeal to two less known foundational approaches (due to Le~Cam and Whittle), and by some additional observations.

The approach of Whittle \cite{Whittle:1970,Whittle:2000} to the foundations of probability theory is based on two independent notions: probabilistic expectations \textit{and} conditional expectations. Both are defined in terms of the Daniell--Stone integrals on the Daniell--Stone vector lattices. For a given set $\X$ the \textit{Daniell--Stone vector lattice} is defined as a lattice $\stone$ with a unit $\II$ such that:
\begin{enumerate}
\item[(i)] $\stone$ is a subset of a set of functions $f:\X\ra\RR\cup\{+\infty\}$, 
\item[(ii)] $f,g\in\stone\limp\sup\{f,g\},\inf\{f,g\}\in\stone$, 
\item[(iii)] $0\leq f\in\stone\limp\inf\{f,\II\}\in\stone$.
\end{enumerate}
The Daniell--Stone \textit{integral} is defined as a map $\omega:\stone\ra\RR$ that is positive, linear, and monotonically sequentially continuous (that is, $\omega(\inf(f_n))=\inf(\omega(f_n))$ for every convergent monotone sequence $\{f_n\}\subseteq\stone$ with $\inf\{f_n\}\in\stone$). The \textit{probabilistic expectation} is defined as a normalised Daniell--Stone integral. The \textit{conditional expectation} is then defined \cite{Whittle:1970} as a function $\E_\omega(\cdot|g)\in\stone$ such that
\begin{equation}        
        \omega((f-\E_\omega(f|g))h(g))=0\;\;\forall f\in\stone\;\forall h:\RR\ra\RR,
\label{def.cond.exp.whittle}
\end{equation}
where $h(g)\in\stone$. For any two probabilistic expectations $\omega$ and $\phi$, the conditional expectation $\E_\phi$ defines uniquely the `updating' (conditioning) of expectation $\omega$ by 
\begin{equation}      
        \omega\mapsto\omega_{\mathrm{new}}:=\omega(\cdot|\E_\phi(\cdot|g)):=\omega\circ\E_\phi(\cdot|g).
\label{exp.updating}
\end{equation}
Given expectations $\omega_1,\omega_2$, we will call $\omega_3(\cdot|\cdot):=\omega_1\circ\E_{\omega_2}(\cdot|\cdot)$ a \textit{conditioned expectation}. The \textit{probability} and \textit{conditional probability} are defined, respectively, by $p(A):=\omega(\chr_A)$ and $p(A|B):=\omega(\chr_A|\chr_B)$, where $\chr_A(x)$ is a characteristic function on $\X$, while $\omega$ is a probabilistic expectation.

The reformulation of probability theory by Whittle refers to the underlying `sample space' $\X$ as a domain of definition of the Daniell--Stone lattice $\stone$. However, it is not necessary to assume this. The Daniell--Stone integration theory requires only to assume that there is given a \textit{Riesz vector lattice}, which is defined by the conditions (ii) and (iii) of definition of Daniell--Stone vector lattice. In such approach, more sophisticated properties of integration theory can be guaranteed by imposing some additional conditions on the Riesz vector lattice. In consequence, the probability theory can be in principle developed for the Riesz vector lattices. This was done by Le~Cam \cite{LeCam:1964,LeCam:1986}, who defined the underlying object of a probability theory as a Riesz vector lattice $\stone$ equipped with a Banach norm, which turns $\stone$ into a Banach lattice.

The Daniell--Stone integration theory on Riesz vector lattices is then a promising candidate for the mathematical framework of information theory built without reference to the notion of probability. However, in order to keep backwards compatibility with the Bayes--Laplace and the Borel--Kolmogorov approaches, as well as in order to keep forwards compatibility with the $W^*$-algebraic reformulation of quantum theory, we need to relate Riesz vector lattices with commutative boolean algebras. This can be provided using the fact that to each boolean algebra $\B$ (which might be finitely additive or countably additive) there corresponds a canonically associated Riesz vector $\stone(\B)$, defined as a set of characteristic function of the open-and-compact subsets of the Stone spectrum of $\B$. The Stone spectrum of $\B$ is defined as a set of all ring homomorphisms from $\B$ to $\ZZ_2$. 

If the underlying boolean algebra $\B$ is countably additive, Dedekind complete, and allows at least one strictly positive semi-finite measure, then it is possible to construct a range of commutative $L_p$ spaces over $\stone(\B)$, with $p\in[1,\infty]$, that do not depend on the choice of the Daniell--Stone integral on $\stone(\B)$ \cite{Fremlin:2000}. We will call such boolean algebras \textit{camDcb-algebras}, and denote the corresponding spaces by $L_p(\B)$. Any given measure space $(\X,\mho(\X),\mu)$ determines an associated camDcb-algebra $\mho$ by
\begin{equation}
        \mho:=\mho(\X)/\{A\in\mho(\X)\mid\mu(A)=0\}.
\label{camDcb.constr}
\end{equation}
On the other hand, accordingly to the Loomis--Sikorski representation theorem \cite{Loomis:1947,Sikorski:1948}, every camDcb-algebra $\mho$ equipped with a measure $\tilde{\mu}$ generates a corresponding measure space $(\X,\mho(\X),\mu)$.

Given a camDcb-algebra $\mho$, we consider the spaces $\M(\mho)\subseteq L_1(\mho)^+$ of finite positive Daniell--Stone integrals over $\mho$ as a kinematic model for commutative information theory.  This way we can unify both standard approaches to foundations of probability theory. The setting of Borel--Kolmogorov approach is recovered by restriction to the evaluational component, via expectation $\omega$ on the lattice of characteristic functions over camDcb-algebra $\mho$ and the Loomis--Sikorski representation of $\mho$ in terms of a measure space for a measure $\tilde{\mu}_\omega$ on $\mho$ defined by this expectation. The setting of Bayes--Laplace approach is recovered by restriction to the relational component, via conditional expectations on the lattice of characteristic functions over finitely additive boolean algebra $\B$.

In order to propose a dynamics of this theory, we will replace the relational component of Whittle's (and Le~Cam's\footnote{The relational component of Le~Cam's approach \cite{LeCam:1964,LeCam:1986} is provided by \textit{transition maps}, which in our case are equivalent with the positive norm-preserving linear maps $T^\coa:\M_1(\mho_1)\ra\M_2(\mho_2)$. The mappings \eqref{exp.updating}, generated by conditional expectations, are special case of these transition maps.}) approach by minimisation of some more general functional on $\M(\mho)$. The reason for the replacement is an important disproportion between the kinematic and dynamic structures of Whittle's approach: while the probabilistic expectation is just a normalised Daniell--Stone integral, the definition of a conditional expectation seems to be an \textit{ad hoc} postulate. If $\omega(f^2)<\infty$, what is equivalent with $f\in L_2(\stone,\omega)$, then \eqref{def.cond.exp.whittle} is equivalent with a variational definition \cite{Whittle:2000}
\begin{equation}
        \E_\omega(f|g):=\arg\inf_{f_e\in\stone}\omega((f-f_e(f,g))^2),
\label{E.var}
\end{equation}
but there is still no justification given of a choice of such functional. Let us note that, due to duality between $L_1(\stone)$ and $L_\infty(\stone)$, the use of conditional expectation as a mapping on the space of integrable functions
\[
        L_\infty(\stone)\ni f\mapsto\E_\phi(f,g)\in L_\infty(\stone)
\]
is equivalent with the use of $\E_\phi$ as a mapping on the space of integrals
\[
        L_1(\stone)\ni\omega\mapsto\omega\circ\E_\phi(\cdot,g)\in L_1(\stone).
\]
This means that conditional expectations can be considered as some particular mappings on information models $\M(\mho)\subseteq L_1(\mho)^+$, which are (dually) defined by some particular variational principle \eqref{E.var}. This leads us to ask about the possibility of introducing other more general dynamic (relational) principle on $\M(\mho)$, which would include the updating by conditional expectations as a special case.
%-------------------------------------------------------------
\subsection{Information geometry}
%-------------------------------------------------------------
Consider the notion of relative distance between the finite positive integrals, defined by the \textit{deviation} (negative \textit{relative entropy}) functionals, 
\[
        D:L_1(\mho)^+\times L_1(\mho)^+\ni(\omega,\phi)\mapsto D(\omega,\phi)\in[0,+\infty],
\]
such that $D(\omega,\phi)\geq0$ and $D(\omega,\phi)=0$ if{}f $\omega=\phi$. Deviation is one of two foundational structures of the information geometry theory \cite{Chentsov:1972,Amari:1985,Morozova:Chentsov:1991,Amari:Nagaoka:1993}, allowing to introduce and analyse the geometric structures on information models $\M(\mho)$. The other is the structure of the differential manifold that can be imposed on $\M(\mho)$ by the local coordinate systems given by embeddings into appropriate Banach spaces (in the finite dimensional case these are $\RR^n$ \cite{Chentsov:1972,Amari:1985}, while in the infinite dimensional case these are $L_{\Phi_1}$ Orlicz spaces \cite{Pistone:Sempi:1995}). We will use the framework of information geometry to define the dynamical principle of the information theory.

Eguchi showed \cite{Eguchi:1983,Eguchi:1985}, for $\dim\M(\mho)<\infty$, that any deviation that is symmetric in first derivatives and has a negative definite hessian determines uniquely (up to a scalar factor) the riemannian metric $g$ and a pair $(\nabla,\nabla^\nsdual)$ of two affine connections on $\M(\mho)$. These objects are defined by the equations
\[
        \begin{array}{c}
        g_\mu(u,v):=(\partial_u)_\mu(\partial_v)_\mu D(\nu,\mu)|_{\nu=\mu},\\
        g_\mu((\nabla_\mu)_u v,w):=-(\partial_u)_\mu(\partial_v)_\mu(\partial_w)_\nu D(\nu,\mu)|_{\nu=\mu},\\
        g_\mu(v,(\nabla^\nsdual_\mu)_u w):=-(\partial_u)_\nu(\partial_w)_\nu(\partial_v)_\mu D(\nu,\mu)|_{\nu=\mu},
        \end{array}
\]
where $(\partial_u)_\mu$ is a G\^{a}teaux derivative at $\mu\in\M(\mho)$ in the direction $u\in\T\M(\mho)$ \cite{Eguchi:1985}. If the torsion and Riemann curvature tensors of both these connections are equal to zero, then $(\M(\mho),g,\nabla,\nabla^\nsdual)$ is called a \textit{dually flat manifold} and there exists a pair $(\ell,\ell^\nsdual)$ of coordinate systems on $\M(\mho)$, called \textit{dually flat coordinates}, such that $\ell$ consist of $\nabla$-geodesics, while $\ell^\nsdual$ consists of $\nabla^\nsdual$-geodesics. They are called \textit{orthogonal} at $q\in\M(\mho)$ if{}f 
\[
        g_q((\partial_\ell)_q,(\partial_{\ell^\nsdual})_q)=0.
\]
This is equivalent to the condition
\[
        \duality{\ell(q),\ell^\nsdual(q)}=0,
\]
where $\duality{\cdot,\cdot}$ denotes the duality between vectors spaces that are codomains of the coordinate systems $\ell$ and $\ell^\nsdual$ (for $\dim\M(\mho)<\infty$ this is just a self-duality of $\RR^n$ understood as vector and covector space, while for $\dim\M(\mho)=\infty$ this is a Banach space duality between Orlicz spaces \cite{Gibilisco:Pistone:1998}). A point $p_\Q\in\Q\subseteq\M(\mho)$ is called a \textit{$\nabla$-projection} of $p\in\M(\mho)$ if{}f the $\nabla$-geodesic connecting $p$ and $p_\Q$ is orthogonal to $\Q$ with respect to $g$, while $\Q\subseteq\M(\mho)$ is called \textit{$\nabla$-convex} if{}f for all $p_1,p_2\in\Q$ there exists a unique $\nabla$-geodesic connecting $p_1$ and $p_2$ and entirely included in $\Q$. As shown by Amari \cite{Amari:1985} (and earlier results by Chentsov \cite{Chentsov:1968,Chentsov:1972} and Csisz\'{a}r \cite{Csiszar:1975}), if $\Q$ is a $\nabla^\nsdual$-convex closed set then the unique $\nabla$-projection $p_\Q$ of $p\in\M(\mho)$ on $\Q$ is equal to \textit{$D$-projection} $\PPP^{D}_\Q(p)$ \cite{Amari:1985}
\begin{equation}
        Q\ni p_\Q=\PPP^{D}_\Q(p):=\arg\inf_{q\in\Q}D(p,q).
\label{D.projection}
\end{equation}
Hence, the dual flatness of the geometry of $\M(\mho)$ is a necessary condition for the existence of the unique $\nabla$-projections on the $\nabla^\nsdual$-convex subspaces provided by the minimum of deviation functional. 
%-------------------------------------------------------------
\subsection{Some special information geometries}
%-------------------------------------------------------------
Let us consider now three important families of deviation functionals (which determine the corresponding families of Norden--Sen geometries).
\begin{enumerate}
\item The family of \textit{Bregman deviations} \cite{Bregman:1967} can be defined for $\dim\M(\mho)<\infty$ by \cite{Bauschke:Borwein:1997,Zhang:2004:divergence}
\[
        D_\Psi(p_1,p_2):=\Psi(\ell_\Psi(p_1))+\Psi^\lfdual(\ell^\nsdual_\Psi(p_2)-\duality{\ell_\Psi(p_1),\ell_\Psi^\nsdual(p_2)},
\]
where $\Psi:\RR^n\ra]-\infty,+\infty]$ is a convex function, $\Psi^\lfdual:\RR^n\ra]-\infty,+\infty]$ is its convex Legendre--Fenchel dual function, defined by 
\[
        \Psi^\lfdual(y):=\sup_{x\in \RR^n}\{\duality{x,y}-\Psi(x)\}\;\;\forall y\in \RR^n,
\]
$\ell_\Psi:\M(\mho)\ra\RR^n$, while its dual function $\ell_\Psi^\nsdual:\M(\mho)\ra\RR^n$ is defined by
\[
        \ell_\Psi^\nsdual(q)=\grad\Psi(\ell_\Psi(q))\;\;\;\forall q\in\M(\mho).
\]
Every Bregman deviation satisfies the \textit{generalised cosine equation} 
\[      
        D_\Psi(p_1,p_2)+D_\Psi(p_2,p_3)-D_\Psi(p_1,p_3)=\duality{\ell_\Psi(p_1)-\ell_\Psi(p_2),\ell_\Psi^\nsdual(p_3)-\ell_\Psi^\nsdual(p_2)}.
\]
Moreover, the Norden--Sen geometry of arbitrary Bregman deviation is dually flat, with dually flat coordinates $(\ell_\Psi,\ell^\nsdual_\Psi)$. Nagaoka and Amari \cite{Nagaoka:Amari:1982} show that any Norden--Sen dually flat geometry with the pair $(\ell,\ell^\nsdual)$ of dually flat coordinates determines a unique corresponding Bregman deviation.
\item The family of \textit{Csisz\'{a}r--Morimoto deviations} \cite{Csiszar:1963,Morimoto:1963,Ali:Silvey:1966} is defined by
\begin{equation}
        D_\fff(\mu,\nu):=\int\mu\,\fff\left(\frac{\nu}{\mu}\right),
\label{Csiszar:Morimoto:dev}
\end{equation}
where $\fff:\RR^+\ra\RR$ is a convex function with $\fff(1)=0$. It was characterised by Csisz\'{a}r \cite{Csiszar:1978} for $\dim\mho<\infty$ by the condition of non-increasing under the information loss provided by partitioning of the space $\X$ and invariance under permutations of partitions. This condition can be restated in representation-independent terms as
\begin{equation}
        D(\mu,\nu)\geq D(\mu\circ T,\nu\circ T),
\label{CS.monotone}
\end{equation}
where $T:L_\infty(\mho)\ra L_\infty(\mho')$ are \textit{Markov maps}, defined as sequentially continuous functions that are linear and positive ($0\leq x\leq\II\limp 0\leq T(x)\leq\II$).\footnote{If the sequential continuity is replaced by a stronger condition of norm continuity or weak topological continuity (which seems to be necessary prerequisite for potential characterisation of \eqref{Csiszar:Morimoto:dev} by \eqref{CS.monotone} in $\dim\mho=\infty$ case), then the maps $T$ are just the duals of Le~Cam's transition maps $T^\coa$ in terms of the Steinhaus--Nikod\'{y}m $L_1(\mho)^\banach\iso L_\infty(\mho)$ Banach space duality.} The condition \eqref{CS.monotone} is called \textit{Markov monotonicity}.
\item The family of \textit{Zhu--Rohwer deviations} \cite{Zhu:Rohwer:1997,Zhu:Rohwer:1998} is defined by 
\[      
        D_\gamma(\mu,\nu):=
        \left\{
        \begin{array}{ll}
        \int\left(\frac{\mu}{1-\gamma}+\frac{\nu}{\gamma}-\frac{\mu^\gamma\nu^{1-\gamma}}{\gamma(1-\gamma)}\right)
        &:\gamma\in\;]0,1[\\
        \int\lim_{\gamma'\ra\gamma}\left(\frac{\mu}{1-\gamma'}+\frac{\nu}{\gamma'}-\frac{\mu^{\gamma'}\nu^{1-\gamma'}}{\gamma'(1-\gamma')}\right)
        &:\gamma\in\;\{0,1\}.
\end{array}
        \right.
\]
It reduces to 
\[
        D_1(\mu,\nu):=\int\left(\mu-\nu+\mu\log\left(\frac{\mu}{\nu}\right)\right)=D_0(\nu,\mu)
\]
for $\gamma\in\{0,1\}$, and to the negative of the \textit{Kullback--Leibler relative entropy}
\[
-\entropy_{KL}(\mu,\nu):=D_1(\mu,\nu)|_{L_1(\mho)_1^+}=\int\mu\log\left(\frac{\mu}{\nu}\right)=D_0(\nu,\mu)|_{L_1(\mho)_1^+}
\]
for $\gamma\in\{0,1\}$ and $L_1(\mho)_1^+:=\{\omega\in L_1(\mho)^+\mid\omega(\II)=1\}$. The family $D_\gamma$ belongs to the class of Bregman deviations. The corresponding dually flat coordinates are given by\textit{ $\gamma$-embeddings} 
\[      
        \ell_\gamma:\M(\mho)\ni\mu\mapsto\frac{\mu^\gamma}{\gamma}\in L_{1/\gamma}(\mho),
\]
which generalise the Nagaoka--Amari \cite{Nagaoka:Amari:1982,Amari:1985} $\gamma$-embeddings and the Zhu--Rohwer \cite{Zhu:Rohwer:1998,Zhu:1998:lebesgue} $\gamma$-embeddings. The Chentsov--Amari dual coordinate system for $\gamma\in]0,1[$ is given by $(\ell_\gamma)^\nsdual=\ell_{1-\gamma}$ (the case $\gamma\in\{0,1\}$ requires separate treatment).
\end{enumerate}
Csisz\'{a}r \cite{Csiszar:1991} showed that $D_1(\mu,\nu)|_{L_1(\mho)_1^+}$ are the unique deviations on $L_1(\mho)_1^+$ that are Csisz\'{a}r--Morimoto deviations \textit{and} Bregman deviations. Amari \cite{Amari:2009:alpha:divergence} showed that $D_\gamma$ are the unique deviations on $L_1(\mho)^+$ that are satisfying this condition. Both proofs are provided for $\dim\mho<\infty$ case.
%-------------------------------------------------------------
\subsection{Information dynamics}
%-------------------------------------------------------------
Using this characterisation od $D_\gamma$, we propose a new approach unifying probability theory and statistical inference theory in a single theory of inductive inference (information dynamics), independent of the notion of probability. Its evaluational-kinematic component was already specified as given by the spaces $\M(\mho)$ of finite positive integrals and their geometry.

Its relational component is given in the following way. Let $P$ be a positive measure on $\M(\mho)$. Define a \textit{$(D,P)$-optimal estimate} with respect to $\Q\subseteq\M(\mho)$ as 
\[
        \PPP^{D,P}(p):=\arg\inf_{q\in\Q}\int_{p\in\M(\mho)}P(p)D(p,q)\in\Q,
\]
and define a \textit{$(D,P)$-ideal estimate} as a $(D,P)$-optimal estimate with respect to $\M(\mho)$. These notions provide the generalisation of the notion of $D$-projection \eqref{D.projection}. Under two assumptions:
\begin{enumerate}
\item[(1)] the inference procedure should select $(D,P)$-optimal estimates,
\item[(2)] $D$ should be Markov monotone Bregman deviation
\end{enumerate}
we can define the \textit{information dynamics} as given by the \textit{constrained maximum relative $\gamma$-entropy updating rule},
\begin{equation}
        \M(\mho)\ni p\mapsto \PPP^{D_\gamma,E}_F(p):=\arg\inf_{q\in\M(\mho)}\left\{\int_{p'\in\M(\mho)}E(p',p)D_\gamma(p',q)+F(q)\right\}\in\M(\mho),
\label{comm.mre.upd}
\end{equation}
where $E(\cdot,p)$ is a positive measure \textit{on} $\M(\mho)$, while $F:\Q\ra]-\infty,+\infty]$ is a constraining function. The map \eqref{comm.mre.upd} provides a selection of $D_\gamma$-optimal estimates with respect to $\Q$ that are weighted by $E$ and satisfy the constraints $F$. If $F(q)$ and $\int_{p'\in\M(\mho)}E(p',p)D_\gamma(p',q)$ are lower semi-continuous and convex in $q$, and if the infimum in \eqref{comm.mre.upd} is finite, then the updating procedure selects a unique $\PPP^{E,D_\gamma}_F(p)$. In particular, $\int_{p'\in\M(\mho)}E(p',p)D_\gamma(p',q)$ is lower semi-continuous and convex, if $E(p',p)=dp'\delta(p-p')$, and in such case the existence and uniqueness of $\PPP^{D_\gamma}_F(p)$ is guaranteed by the requirement that $\ell_{1-\gamma}(\Q)$ is non-empty, closed and convex (see e.g. \cite{Jencova:2005}). The measure $E(\cdot,p)$ will be called a \textit{prior relative to $p$}.

We consider (\ref{comm.mre.upd}) as a fundamental principle of  information dynamics (in commutative case). In our opinion, its status in information theory is analogous to the status of lagrangean variational principle in classical field theory.

The temporal character of this information dynamics might enter through the dependence of constraints $F$ on the external time parameter $t$. If $E(p',p)=dp'\delta(p-p')$, then $p$ plays a role of the initial state of this dynamics. Under these conditions, if the constraints $F(q)$ have the form $F(q,t)$, where the parameter $t$ is interpreted as `time', if $F(q,t=t_0)=0$, and if the solutions $\PPP^{D_\gamma}_F(p)$ also depend on $t$, then the trajectory
\[
        p(t):=\PPP^{D_\gamma}_{F(t)}(p_0)\in\M(\mho)
\]
with $p_0:=\PPP^{D_\gamma}_{F(t=t_0)}(p)$ can be considered as a temporal evolution generated by the constraint $F(q,t)$ of relative entropic information dynamics \eqref{comm.mre.upd}. We will call $p(t)$ \textit{temporal information dynamics} or \textit{entropic evolution of information}.
%where $E(\cdot,p)$ is a positive measure \textit{on} $\M(\mho)$, while $F:\Q\ra]-\infty,+\infty]$ is a constraining function. This map provides a selection of $\gamma$-optimal estimates with respect to $\Q$ that are weighted by $E$ and satisfy the constraints $F$. If $F$ is lower semi-continuous, if $\ell_\gamma(\Q)$ is non-empty, closed and convex, if $E(p',p)=dp'\delta(p-p')$, and if the infimum in \eqref{comm.mre.upd} is finite, then the updating procedure selects a unique $p_{\mathrm{new}}$. Under these conditions, if the constraints $F(q)$ depend on some parameter $t$ that is interpreted as `time', if $F(q,t=t_0)=0$, and if the solutions $p_{\mathrm{new}}$ also depend on $t$, then the trajectory $p_{\mathrm{new}}(t)\in\M(\mho)$ can be called a temporal evolution generated by the constraint $F(q,t)$ of relative entropic information dynamics \eqref{comm.mre.upd}.
%-------------------------------------------------------------
\subsection{Reconstruction of the standard frameworks}
%-------------------------------------------------------------
Our approach provides a particular solution to the problem of the relationship between information, inference and probability. The probability theory is just a special case of the information theory and requires no additional justification. The Whittle approach is recovered by the fact \cite{BGW:2005} that the duals of conditional expectations are characterised as minimisers of expectations of Bregman deviations, the Borel--Kolmogorov approach is recovered by forgetting the relational component, and passing to probabilistic measures, while the Bayes--Laplace approach is recovered by normalisation and by the fact \cite{Caticha:Giffin:2006,Giffin:2009} that the Bayes rule is just a special case of the entropic updating rule with Dirac's delta constraints. 

Note that the differences between $\gamma$-ideal estimates (defined as $(D_\gamma,dp'\delta(p'-p))$-ideal estimates) appear only in the third order of the Taylor series expansion of $D_\gamma$. Up to the second order, all $D_\gamma$ are determined by the same riemannian metric $g$, which reduces to the Fisher--Rao metric for $L_1(\mho)_1^+$. Hence, the inferences based on $D_\gamma$ on $\M(\mho)$ agree with the inferences based on $D_{1/2}$, which in turn agree with the inferences based on the Hilbert space $L_2$ norm. However, beyond the second order, the use of the $L_2$ space and its norm is \textit{not} justified. Moreover, the projections in Hilbert space are linear operators, hence they can deal only with linear problems (constraints), as opposed to non-linear $\gamma$-projections onto convex subspaces of $L_{1/\gamma}$ spaces (with $L_1$ spaces playing most important role). The constrained relative $\gamma$-entropy updating incorporates Hilbert space based methods of projection and estimation as a special case. Under restriction to $L_1(\mho)_1^+$, the only justified method of inference is given by constrained maximisation of the Kullback--Leibler relative entropy $\entropy_{KL}$.
%-------------------------------------------------------------
%\subsection{Interpretational remarks}
%-------------------------------------------------------------
%-------------------------------------------------------------
\section{Towards new foundations of quantum theory}
%-------------------------------------------------------------
In this section, we will present the elements of a new approach to foundations of quantum theory, which is a \textit{direct} generalisation of our reformulation of foundations of probability theory and statistical inference theory. The steps of the construction of this approach are the following: (1) construction of quantum information kinematics based on integration theory on non-commutative algebras and quantum information geometry of the spaces of the positive finite integrals, (2) construction of quantum information dynamics based on maximisation of constrained quantum relative entropy, (3) reconstruction of the orthodox kinematical and dynamical Hilbert space based framework for quantum theory as a special case of a new framework, (4) construction of adequate interpretation that covers the conceptual issues of the quantum theory using the new setup. We will discuss here mainly the points (1) and (2).
%-------------------------------------------------------------
\subsection{Non-commutative integration theory}
%-------------------------------------------------------------
An algebraic approach to the mathematical framework of quantum theory \cite{Segal:1947:postulates,Segal:1953} replaces the consideration of abstract Hilbert space by the consideration of abstract \textit{$C^*$-algebra} $\A$ with unit $\II$, which is also a Banach space over $\CC$, and is  equipped with a map $^*:\A\ra\A$ such that 
\[
        (ab)^*=b^*a^*,\;\;\;(a+b)^*=a^*+b^*,\;\;\;(\lambda a)^*=\bar{\lambda}a^*,\;\;\;(a^*)^*=a,
\]
\[
        \n{a^*}=\n{a},\;\;\;\n{ab}\leq\n{a}\n{b},\;\;\;\n{\II}=1,\;\;\;\n{a^*a}=\n{a}^2.
\]
An \textit{algebraic state} is a functional $\omega:\A\ra\CC$ that is linear, positive and normalised. A basic example of a $C^*$-algebra is $\BH$, while a basic example of an algebraic state on it is $\tr(\rho\cdot):\BH\ra\CC$, where $\rho$ is a density operator. While every algebraic state $\omega$ on $\BH$ can be represented in terms of $\tr(\rho_\omega\cdot)$, it is no longer true for other $C^*$-algebras that naturally arise in relativistic quantum field theory and continuous quantum statistical mechanics (see e.g. \cite{Bratelli:Robinson:1979,Haag:1992}).

For any pair of a $C^*$-algebra $\A$ and linear, positive functional $\omega$ on $\A$, the Gel'fand--Na\u{\i}mark--Segal (GNS) theorem associates a unique Hilbert space $\H_\omega$ and a unique (up to unitary equivalence) representation $\pi_\omega:\A\ra\BBB(\H_\omega)$ such that there exists $\Omega_\omega\in\H_\omega$ that is cyclic for $\pi_\omega(\A)$, $\n{\Omega_\omega}^2=1$, and $\omega(A)=\s{\Omega_\omega,\pi_\omega(A)\Omega_\omega}_\omega$ for all $A\in\A$. This theorem allows to bypass the mathematical setting of Hilbert spaces, dealing directly with the representation-independent properties of quantum theoretic formalism in terms of $C^*$-algebras and linear positive functionals on them.

In order to construct the new formulation of quantum theory, we need to resign from Hilbert spaces as the kinematic setting \textit{and} to replace it by the setting based on integration theory on non-commutative algebras. Let us start from identification of the correct notions of an `integrable non-commutative algebra' and of an `integral' on it. A \textit{$W^*$-algebra} is defined as such $C^*$-algebra $\N$ which is a Banach dual to some Banach space. This Banach space is called a \textit{predual} of $\N$ and is denoted by $\N_*$. The $W^*$-algebra generalises the notion of an `integrable' commutative algebra to the non-commutative case. The space $\N_*$ consists of all functionals on a $W^*$-algebra $\N$ that are \textit{normal}, that is, $\omega(\sup\filter)=\sup_{a\in\filter}\omega(a)$ for every directed filter $\filter\subseteq\N$ with the upper bound $\sup\filter$. The spaces $\N_*^+=:\{\omega\in\N_*\mid\omega(A^*A)\geq0\}$ of \textit{quantum information states} provide the non-commutative analogue of the spaces $\M(\mho)$ of finite Daniell--Stone integrals on $\stone(\mho)$. The spaces $\N^+_{*1}:=\{\omega\in\N_*^+\mid\omega(\II)=1\}$ of normalised quantum information states provide the non-commutative analogue of the spaces $L_1(\mho)_1^+$. 

This is actually much more than just analogy. Falcone and Takesaki \cite{Falcone:Takesaki:2001} formulated the canonical theory of non-commutative $L_p(\N)$. The elements of $L_p(\N)$ can be represented in form $x\omega^{1/p}$, where $x\in\N$ and $\omega\in\N_*$. This theory provides a notion of the integral $\int:L_1(\N)\ra\CC$ such that, in particular, $\int x\omega=\omega(x)\in\CC$. This integral is meaningful also for such formulas like 
\[
        \int x_1\phi_1^{\alpha_1}\cdots x_n\phi_n^{\alpha_n}
\]
for $\phi_1,\ldots,\phi_n\in\N_*^+$ and $x_1,\ldots,x_n\in\N$ and $\alpha_1,\ldots,\alpha_n\in\{z\in\CC\mid\re(z)\geq0\}$ with $\sum_{i=1}^n\alpha_i=1$. The bilinear map 
\[
        L_1(\N)\times\N\ni(x,T)\mapsto\duality{x,T}:=\int xT\in\CC
\]
gives the pairing between $\N$ and $L_1(\N)$ which identifies $L_1(\N)$ with $\N_*$. By means of definition, $L_\infty(\N)$ is identified with $\N$. The duality between $L_p(\N)$ and $L_q(\N)$, for $1/p+1/q=1$, is given by 
\[
        L_p(\N)\times L_q(\N)\ni(S,T)\mapsto\duality{S,T}=\int ST\in\CC.
\]
The analytic structure of $L_p(\N)$ is defined by the Banach norm $\n{T}_p:=(\int|T|^p)^{1/p}$.

From the perspective of non-commutative integration theory, evaluation of $\tr(\rho\cdot)$ on $\BH$ as well as the evaluation of probability measure $\tr(\rho\EE(\cdot))$ on boolean algebra $\mho(\X)$ (where $\EE$ is a semi-spectral measure, while $\X$ is usually identified with the spectrum of some operator) can be replaced by acting with $\omega\in\N_{*}^+\iso L_1(\N)^+$ on $W^*$-algebra $\N$. Because semi-spectral measures $\EE$ are no longer used in this setting in order to define the probabilistic evaluations,  von Neumann's spectral theorem loses its importance. Hence, the associated problems of contextuality of definition of operators corresponding to probabilistic descriptions, particular quantitative experimental setups and results of their use (see e.g. \cite{Busch:Grabowski:Lahti:1995}) disappear. The \textit{only} meaning of elements of $W^*$-algebras in our approach is: they are the \textit{statements} of an (intersubjectively shared) \textit{abstract language} subjected to quantitative evaluation. 

Now we need to equip this general setting with the information geometric and dynamical structures. 
%-------------------------------------------------------------
\subsection{Quantum information geometry}
%-------------------------------------------------------------
Our novel result is a generalisation of quantum information geometric structures which is based the Falcone--Takesaki theory. In analogy with Nagaoka--Amari \cite{Nagaoka:Amari:1982,Amari:Nagaoka:1993} \textit{and} Zhu--Rohwer \cite{Zhu:Rohwer:1997,Zhu:Rohwer:1998} approaches, we define the \textit{quantum $\gamma$-embeddings} ($\gamma$-coordinates) by
\[
        \ell_\gamma:\N_*^+\ni\omega\mapsto\ell_\gamma(\omega):=\omega^\gamma/\gamma\in L_{1/\gamma}(\N).
\]
We define the \textit{quantum $\gamma$-deviation} $D_\gamma:\N_*^+\times\N_*^+\ni(\omega,\phi)\mapsto D_\gamma(\omega,\phi)\in\RR$ by
\[
        D_\gamma(\omega,\phi):=\int\left(\frac{\omega}{1-\gamma}+\frac{\phi}{\gamma}-\re(\ell_\gamma(\omega)\ell_{1-\gamma}(\phi))\right)=\int\left(\frac{\omega}{1-\gamma}+\frac{\phi}{\gamma}-\frac{\re(\omega^\gamma\phi^{1-\gamma})}{\gamma(1-\gamma)}\right),
\]
for $\gamma\in]0,1[$, and by the limit $\gamma\ra\gamma'$ under integral sign for $\gamma'\in\{0,1\}$. The \textit{quantum relative $\gamma$-entropy} is defined as $\entropy_\gamma(\omega,\phi):=-D_\gamma(\omega,\phi)$, and $-\entropy(\phi,\omega):=D_0(\omega,\phi)=D_1(\phi,\omega)$. This definition reduces to the Jen\v{c}ov\'{a}--Ojima \cite{Jencova:2003:arXiv,Jencova:2005,Ojima:2004} quantum $\gamma$-deviation for any particular choice of reference weight $\psi$ (due to isometric isomorphism of the Falcone--Takesaki $L_p(\N)$ spaces with the Kosaki--Terp $L_p(\N,\phi)$ spaces), to the Umegaki--Araki quantum relative entropy \cite{Umegaki:1962,Araki:1976:relative:entropy:I,Araki:1977:relative:entropy:II} in the Petz \cite{Petz:1986:properties} reference-independent form $\lim_{t\ra+0}\frac{i}{t}\phi([\DD\omega:\DD\phi]_t-\II)$ for $\omega,\phi\in\N_{*1}^+$, to the Hasegawa quantum $\gamma$-deviation \cite{Hasegawa:1993} $\tr(\rho_\phi-\rho_\omega^\gamma\rho_\phi^{1-\gamma})/(\gamma-\gamma^2)$ for $\omega=\tr(\rho_\omega\cdot)$, $\phi=\tr(\rho_\phi\cdot)$ and type I $W^*$-algebra $\N$, and to the Zhu--Rohwer $\gamma$-deviation $D_\gamma(\omega,\phi)$ for commutative $\N$. The symbol $[\DD\omega:\DD\phi]_t$ denotes Connes' cocycle \cite{Connes:1973:classification}, which is a non-commutative generalisation of the Radon--Nikod\'{y}m derivative.

The properties of $D_\gamma$ are analysed in \cite{Kostecki:2011:OSID}. They are the same as the properties of Jen\v{c}ov\'{a}--Ojima deviation \cite{Jencova:2005}, but with a key difference that the definition and properties of $D_\gamma$ explicitly do not depend on any reference weight $\psi$. For $\gamma\in]0,1[$, $D_\gamma$ is both Petz's $\fff$-deviation \cite{Petz:1985:quasientropies} (which is a non-commutative generalisation of the Csisz\'{a}r--Morimoto deviation) and the non-commutative generalisation of Bregman's deviation (see \cite{Jencova:2005,Petz:2007:Bregman} for special cases, and \cite{Kostecki:2011:OSID} for a general definition). $D_\gamma$ (resp., $\entropy$) is convex and lower semi-continuous on the space $\N_*^+\times\N_{*0}^+$ (resp., $\N_*^+\times\N_*^+$) endowed with the product of norm topologies \cite{Jencova:2005} (resp., the topology of pointwise convergence on $\N_*$ \cite{Ohya:Petz:1993}). Here $\N_{*0}^+:=\{\omega\in\N_*^+\mid\omega(x^*x)=0\limp x=0\}$ denotes the space of all \textit{faithful} quantum information states on $\N$. If $\Q\subseteq\M(\N)$ is weakly closed and if $\ell_\gamma(\Q)$ is convex, then there exists a unique $D_\gamma$-projection on $\Q$,
\[
\M(\N)\ni\omega\mapsto\PPP_\Q^{D_{1-\gamma}}(\omega)=\arg\inf_{\phi\in\Q}D_{1-\gamma}(\omega,\phi)\in\Q.
\]

Following Amari's and Csisz\'{a}r's results in commutative case, \textit{we conjecture that $D_\gamma$ (resp., $D_1$) is a unique deviation on $\N_*^+$ (resp., $\N^+_{*1}$) that belongs to both Petz's and non-commutative Bregman's families of quantum deviations}.

In \cite{Jencova:2006,Jencova:2010} Jen\v{c}ov\'{a} has introduced the differential manifold structure on the space $\N_{*0}^+$. Using this differential structure, one can consider the tangent spaces, riemannian metrics and affine connections on $\M(\N)\subseteq\N_{*0}^+$. For $\dim\M(\N)=\infty$, the construction of tangent space of $\M(\N)$ is a delicate issue. As a result, the standard approach \cite{Gibilisco:Pistone:1998,Gibilisco:Isola:1999} is to define differential geometric objects on $\M(\N)$ using the family of vector $\gamma$-bundles $\F^\gamma$ over $\M(\N)$ instead of tangent bundle $\T\M(\N)$. Following Gibilisco--Pistone \cite{Gibilisco:Pistone:1998}, Gibilisco--Isola \cite{Gibilisco:Isola:1999} and Jen\v{c}ov\'{a} \cite{Jencova:2005}, we generalise the \textit{$\gamma$-representation} of $\gamma$-bundles using the embeddings into the tangent spaces of unit spheres $S_{1/\gamma}(\N)$ of non-commutative $L_{1/\gamma}(\N)$ spaces,
\[
        \F^\gamma_\omega(\M(\N))\ni v\mapsto\ell^{\#(\omega)}_\gamma(v):=x\phi^\gamma\in T_{\omega^\gamma}S_{1/\gamma}(\N),
\]
where 
\begin{align}
        T_\omega S_{1/\gamma}(\N):&=\{x\phi^\gamma\in L_{1/\gamma}(\N)\mid\re\int\omega^{1-\gamma}x\phi^\gamma=0\}\notag\\
        &=\{x\phi^\gamma\in L_{1/\gamma}(\N)\mid\re\omega([\DD\omega:\DD\phi]_{i\gamma}(x))=0\}.
\end{align}
In the special cases, this definition reduces to those given in \cite{Jencova:2005,Gibilisco:Isola:1999}, and \cite{Gibilisco:Pistone:1998}. It allows us to define a \textit{quantum $\gamma$-metric} $g^\gamma_\omega:\F^\gamma_\omega(\M(\N))\times\F^\gamma_\omega(\M(\N))\ra\RR$ by
\begin{equation}
        g^\gamma_\omega(u,v):=\re\int\ell^{\#(\omega)}_\gamma(u)(j_\gamma\circ\ell_\gamma)^{\#(\omega)}(v)=\re\int\omega^{1-\gamma}x\omega^\gamma y,
\label{gamma.metric}
\end{equation}
where $j_\gamma:=\ell_{1-\gamma}\circ\ell_{\gamma}^{-1}$ and $x,y\in\N$. The quantum $\gamma$-metric \eqref{gamma.metric} generalises the Wigner--Yanase--Dyson $\gamma$-metric \cite{Wigner:Yanase:1963,Hasegawa:1993}. The generalisation of the Gibilisco--Isola family of $\gamma$-connections \cite{Gibilisco:Isola:1999}, based on the projection of trivial affine structure of non-commutative $L_{1/\gamma}(\N)$ spaces onto unit spheres $S_{1/\gamma}(\N)$ is straightforward, and will be discussed, together with other quantum information geometric structures, in \cite{Kostecki:2011:QIG}.% and quantum $\gamma$-connections of \cite{Gibilisco:Isola:1999}, respectively.
%-------------------------------------------------------------
\subsection{Quantum information dynamics}
%-------------------------------------------------------------
Using Bauer's maximisation principle \cite{Bauer:1958} and the results of Jen\v{c}ov\'{a} \cite{Jencova:2005}, we conclude that if $\Q\subseteq\M(\N)$ is a nonempty set such that $\ell_\gamma(\Q)\subset L_{1/\gamma}(\N)$ is weakly closed and convex, and $F:\Q\ra]-\infty,+\infty]$ is a weakly-* lower semi-continuous convex function, then there exists a unique 
\[
\PPP^{D_\gamma}_F(\omega):=\arg\inf_{\phi\in\M(\N)}\{D_{1-\gamma}(\omega,\phi)+F(\phi)\}
\]
if this infimum is finite. 

We define the \textit{quantum constrained maximum relative $\gamma$-entropy updating rule} by
\begin{equation}        \M(\N)\ni\omega\mapsto\PPP^{D_\gamma,E}_F(\omega):=\arg\inf_{\phi\in\M(\N)}\left\{\int_{\varphi\in\M(\N)}E(\varphi,\omega)D_\gamma(\varphi,\phi)+F(\phi)\right\}\in\M(\N),
\label{maxentq}
\end{equation}
for the positive measure $E(\cdot,\omega)$ on $\M(\N)$, and constraints given by convex weak-* lower semi-continuous $F:\Q\ra]-\infty,+\infty]$ with convex closed $\ell_{1-\gamma}(\Q)$, which guarantees the existence and uniqueness of the result of this updating when this infimum takes a finite value and if $E(\varphi,\omega)=d\varphi\delta(\varphi-\omega)$. This rule equips the evaluational theory of integrals on non-commutative algebras with the relational (dynamical) counterpart, which provides a selection of $\gamma$-optimal estimates that are weighted by $E$, relative to $\Q$, and constrained by $F$. This generalises and replaces the Hilbert space based approach. The convex closed subspaces of Hilbert spaces associated with the orthogonal linear projection operators are replaced in our approach by the convex closed subspaces of non-commutative $L_{1/\gamma}(\N)$ spaces, which used as the codomain of quantum information geometric embeddings $\ell_\gamma$ of the functionals $\omega\in\M(\N)$. If $E(\varphi,\omega)=d\varphi\delta(\varphi-\omega)$, the constraints $F(\phi)$ depend on (discrete or continuous) `time' parameter $t$ in a way that $F(\phi,t=t_0)=0$ and the solutions of \eqref{maxentq} reflect this dependence on $t$, then the resulting trajectory $\PPP^{D_\gamma}_{F(t)}(\omega_0)$ on quantum information manifold, where $\omega_0:=\PPP^{D_\gamma}_{F(t=t_0)}(\omega)$ can be understood as a non-linear \textit{temporal dynamics of quantum states}.

Interpretation of spaces $\M(\N)$ as a general framework for non-linear quantum theoretic models and interpretation of the  updating rule \eqref{maxentq} as a general description of their non-linear behaviour (quantum dynamics) opens new perspectives for the foundations and applications of quantum theory. %If Petz's deviation and non-commutative Bregman deviation can be equipped with clear infomration theoretic characterisatin, and if our conjecture holds, then the choice of deviation $D_\gamma$ in dynamical rule \eqref{maxentq} would possess justification based on clear information theoretic postulates, providing a concrete reply to requests of \cite{Fuchs:2003} and \cite{Clifton:Bub:Halvorson:2003}. 
%-------------------------------------------------------------
\subsection{Reconstruction of the Hilbert space based framework}
%-------------------------------------------------------------
The reconstruction of kinematic setting of ordinary Hilbert space based quantum theory is provided by the quantum information geometric representation of $\M(\N)$ in terms of the non-commutative $L_2(\N)$ space,
\[
        \ell_{1/2}:\M(\N)\ni\phi\mapsto 2\phi^{1/2}\in L_2(\N).
\]
The space $L_2(\N)$ equipped with the inner product 
\[
        L_2(\N)\times L_2(\N)\ni(S,T)\mapsto\s{S,T}:=\int T^*S\in\CC
\]
is a Hilbert space that is unitary isomorphic with the Haagerup representation Hilbert space $\H_H$ \cite{Haagerup:1975:standard:form}, and is unitary isomorphic with the GNS Hilbert space $\H_\phi$ for any choice of $\phi\in\N_{*0}^+$. The corresponding $1/2$-deviation takes the form 
\[
        D_{1/2}(x,y)=\frac{1}{2}\n{x-y}_{\H_H}^2.
\]
The $D_{1/2}$-projection is a minimiser of the Hilbert space norm, and defines a unique corresponding orthogonal projection operator. Hence, it is possible to recover the Hilbert space geometry as a special case of quantum information geometry. 

In \cite{Kostecki:2011:towards:series} we will present the derivation of temporal behaviour of ordinary Hilbert space based quantum theoretic models from the relative entropic information dynamics on $\M(\N)$. The brief idea of this recovery is following. The entropic updating generates non-linear mappings of Hilbert space vectors or density operators (that is often called a ``reduction due to measurement''). In particular case, it reduces to the von Neumann--L\"{u}ders rule. However, in addition, to each stage of the non-unitary entropic evolution, there is associated a unique unitary evolution, specified by the Tomita--Takesaki modular automorphism, which generates the unitary evolution of vectors that represent the quantum information states of $\M(\N)$ within $\H_H$. Because the Tomita--Takesaki automorphism is completely determined by underlying quantum information state, the entropic updating leads also to recovery of perturbations of generators of unitary dynamics that are determined by the constraints of dynamics (that is often called a ``perturbation due to interaction'').
%-------------------------------------------------------------
\section{Some remarks on interpretation}
%-------------------------------------------------------------
We will finish this paper with a few remarks on the interpretation that we associate with the above formalism. These remarks are not intended to be comprehensive. Their role is only to indicate some of the key ideas. The extensive discussion of the conceptual issues related with our approach will be provided in \cite{Kostecki:2011:towards:series}. See also \cite{Kostecki:2011:principles} for a further discussion of information semantics and intersubjective interpretation in a context of foundations of probability theory and statistical inference.
%-------------------------------------------------------------
\subsection{Inductive inference and information semantics}
%-------------------------------------------------------------
In general, by \textit{inductive inference} we understand some form of inductive logical reasoning, as opposed to deductive logical reasoning. The latter specifies premises by the valuations of sentences in \textit{truth} values, and provides an inference procedure which is considered to lead to \textit{certain} conclusion on the base of given premises. The former specifies premises by the valuations of sentences in \textit{possible} (\textit{plausible}) values and provides an inference procedure which is considered to lead to \textit{most possible} (\textit{most plausible}) conclusions on the base of given premises. From the mathematical perspective, the difference between deductive and inductive inference lays not in the form of logical valuations (these can be the same in both methods), but in the procedure of specifying conclusions on the base of premises. The conclusions of multiple application of deductive inference to the sequence of sets of premises depend \textit{in principle} on all elements of all these sets, while the conclusions of the multiple application of inductive inference to the sequence of sets of premises depend \textit{in principle} only on some elements of some of these sets. By this reason, the premises of inductive inference are also called \textit{evidence}. An example of inductive inference procedure is any statistical reasoning based on probabilities. The evidence (called also `constraints of inference') can consist, for example, of particular quantities with units called `experimental data' \textit{together with} a particular choice of a method which incorporates these `data' into statistical inference. 
Any choice of such method \textit{defines} the actual meaning of the `data', and is a crucial element of the inference procedure. A standard example of such method is to ignore everything what is known about a sequence of numbers from a series of repeated measurements associated with a single abstract quality (such as a ``position''), leaving only the value of arithmetic average and the value of a fluctuation around this average as a subject of comparison (e.g., by identification) with the mean and variance parameters of the gaussian probabilistic model.

Let us now consider the mathematical framework briefly presented in previous sections. On the level of \textit{information semantics}, the underlying algebra ($\mho$ or $\N$) represents an \textit{abstract qualitative language} subjected to quantitative evaluation, the spaces ($\M(\mho)$ or $\M(\N)$) of finite integrals and their geometry (kinematic models of information) represent \textit{quantified knowledge}, while the entropic updating (dynamics of information models) represents \textit{quantitative inductive inference}. 

This means that the algebra $\N$ is \textit{not} considered as consisting of `observables' or `elementary events'. Due to its kinematical character, the quantum information model $\M(\N)$ and its particular geometry is considered as a model of \textit{potential} knowledge. In this sense, it would be much more relevant to use the word `observable quantity' in order to refer to elements of $\M(\N)$ than to the elements of $\N$. (This is especially appealing when one notices that orthodox quantum theoretical approach identification of self-adjoint elements of non-commutative algebra with `observables' is applied only to \textit{concrete} algebras---such that act on the Hilbert spaces---and not to the abstract algebras. However, the passgage from abstract to concrete algebra, $\pi_\omega:\N\ra\BBB(\H_\omega)$ involves the dependence on $\omega\in\M(\N)$.) Suggestions of terminological shift of this type are present e.g. in \cite{Klauder:1997,Neumaier:2003}, but such change would create an unnecessary semantic confusion, hence we prefer to simply resign from using the term `observable'. 

The quantitative information dynamics of the model $\M(\N)$ is formed by the additional choices of $D_\gamma$, $E$ and $F$. On the semantic level, the functions $E$ and $F$ specify the \textit{evidence} subjected to the inductive inference rule provided by entropic information dynamics \eqref{maxentq}. The resulting projection $\PPP^{D_\gamma,E}_F$ is an inference: specification of most plausible state of knowledge subjected to given evidence. For a `temporal history' $F=F(t)$ and an `initial state' $\omega$ specified by $E(\varphi,\omega)=d\varphi\delta(\varphi-\omega)$, the information dynamics \eqref{maxentq} takes a form of temporal evolution of quantum states $\omega(t):=\PPP^{D_\gamma}_{F(t)}(\omega_0)$. It models the changes of the \textit{actual} knowledge determined by the changes of what is considered to be an \textit{actual} evidence.

This way quantum theory becomes a theory of quantitative inductive inference (information dynamics) provided with respect to the qualitative abstract language of non-commutative algebras. The constraints of this inductive logic are imposed by the choice of particular $\M(\N)$, $D_\gamma$, $E$ and $F$. 
%---------------------------------------------------------------------
\subsection{Intersubjective interpretation and farewell to ontology}
%---------------------------------------------------------------------
The above information \textit{semantics} requires an additional \textit{interpretation} which would determine the particular operational and conceptual meaning attributed to the terms `knowledge' and `change of knowledge'. In particular, this interpretation should determine the choice of a particular information kinematics (that is, $\M(\N)$ and its information geometry) and a particular information dynamics ($D_\gamma$, $E$, $F$) when applied to some particular experimental situations. 

In order to deal with the conceptual problems of quantum theory and to bypass the conflict between `objective' and `personalistic' bayesian interpretations of probability and statistical inference theory, we propose new \textit{intersubjective interpretation}. According to it, the knowledge used to define particular theoretical model should bijectively correspond to the knowledge that is sufficient and necessary in order to \textit{intersubjectively reproduce} corresponding `experiment of a given type' that is intended to be described by this theoretical model. An experiment of a given type consists of an experimental setup of a given type and its particular use, which amounts to setting a particular configuration (of controlled actual inputs) and a particular registration scale of potential outcomes (allowed results of use). An `experiment of a given type' is an idealistic abstraction. However, this abstraction needs not to be understood in ontological sense. We consider it as a purely epistemic entity. The crucial question is: how to verify whether some particular individual setup under consideration and some particular actions and observations associated with it can be considered as an \textit{intersubjectively valid} instance of an experiment of a given type? The answer is: an agreement with some particular knowledge has to be positively verified in operational terms. Hence, this knowledge actually \textit{defines} an intersubjective notion of an experiment of a given type. The intersubjective interpretation amounts to require the bijective agreement between the kinematics of theoretical model and the terms of experimental setup construction, as well as the bijective agreement between the dynamics of theoretical model and the terms of use of experimental setup. We postulate bijection and not identification because we allow complete separation between the theoretical abstract language used to intersubjectively define and communicate theoretical models, and the operational language used to intersubjectively define and communicate corresponding experiments. In consequence, the intersubjective interpretation does not define the absolute (passive, static) meaning of the notion of `knowledge'. It defines only the relational (active, dynamic) meaning of this notion, as a particular relationship between kinematics-and-dynamics of theoretical model and construction-and-use of experimental setup.\footnote{By the same reason (intersubjective context dependent bijective correspondence as opposed to absolutist identification), intersubjective interpretation cannot be considered as operationalism. On the other hand, it provides no ontological claims. Thus, it bypasses the na\"{\i}vet\'{e}s of conflict between `realism' and `operationalism'.}

This interpretation introduces a key property expected from any scientific theory directly into conceptual foundations of quantum theory: a requirement that theory should allow unambiguous intersubjective verification in an experiment of a particular type. We require that every quantum theoretic model has to be \textit{defined} in terms of some particular experimental design that can be used to intersubjectively verify the predictions of this theoretical model. 

Quite similar postulate was already proposed in the context of semi-spectral approach to foundations of quantum theory \cite{Grabowski:1989,Busch:Grabowski:Lahti:1989,Busch:Grabowski:Lahti:1995,deMuynck:2002}. However, the semi-spectral approach is involved in the mathematical frameworks of Hilbert spaces and measure spaces which play no foundational role in our approach. The basic elements of our new mathematical foundations for quantum theory are interpreted as follows. An algebra $\N$ is understood as an abstract qualitative language used as a common reference in an intersubjective communication about the abstract (idealised, theoretical, intentional) qualities that are subjected to quantification (quantitative evaluation, integration) in the course of the use of experimental setups. The quantum information model $\M(\N)\subseteq\N_*^+$ and its geometry is understood as the carrier of quantitative intersubjective knowledge describing the particular experimental setup under consideration. The choices of evidence $E$ and $F$ provides the description of particular control settings and particular range of allowed response outcomes.  Together with the choice of $D_\gamma$, these choices determine intersubjectively accepted range (scale) of potential of quantitative results of use (outcomes) of experimental setup that is modelled in terms of $\M(\N)$. As a consequence, the temporal information dynamics $\PPP^{D_\gamma}_{F(t)}(\omega_0)$ provides the time-dependent description of most plausible response outcomes that can be inferred from the given evidence.

We do not assume any \textit{a priori} division of the notion of experimental setup and its use into ``measuring device'' and ``measured system''. The \textit{mathematical notion of} a ``measured system'' can be introduced as a special property (decomposition into tensor product) imposed on the algebra, quantum state space and quantum dynamics, providing this way the backwards compatibility of mathematical formalisms and their predictive validity. This way we can reintroduce suitable idealistic concept of a ``measured system'' for \textit{some} experimental situations that allow it. Yet, this concept obtains here purely epistemic meaning, and it has no foundational role. Moreover, it might be just non-applicable in construction of theoretical models that correspond to \textit{some} experimental situations. From the perspective provided by our approach, the \textit{idealistic concept of} a ``measured system'' understood as kantian \textit{Ding an sich} is completely irrelevant both for foundations and applications of quantum theory. Thus, the ontological interpretations of quantum theoretic formalism, as well as the internal problems of these interpretations (that are usually introduced by the use of such operationally undefined idealistic terms as `matter', `fields', `particles', `quantum objects', `randomness', `universe', `nature'), are also completely irrelevant for the foundations of quantum theory.
%------------------------------------------------------------
\section*{Acknowledgments}
%------------------------------------------------------
{\small I am grateful to Chris Isham and Stanis{\l}aw L. Woronowicz for very valuable discussions, to Jerzy Lewandowski for the support of my research, to Ariel Caticha, Paolo Gibilisco, Frank Hellmann, Ewa Infeld, Wojtek Kami\'nski and W{\l}odek Natorf for their comments and suggestions on the early version of this paper, and to Fotini Markopoulou, Olaf Dreyer, Jamie Vicary, Cecilia Flori, Ewa Infeld and Tomasz Ko{\l}odziejski for helping with accomodation during my scientific visits. This work was partially supported by the FNP \textit{Mistrz} grant 2007, ESF \textit{Quantum Geometry and
Quantum Gravity} short visit grant 1955, and ESF \textit{Quantum Geometry and Quantum Gravity} exchange grant 2706.}
%------------------------------------------------------
{\scriptsize
\bibliographystyle{../rpkbib}
\bibliography{../rpkrefs}

\begin{thebibliography}{100}

\bibitem{Ali:Silvey:1966}
{Ali S.M., Silvey S.D.}, 1966, \textit{A general class of coefficients of
  divergence of one distribution from another}, J. Roy. Stat. Soc. B
  \textbf{28}, 131.

\bibitem{Amari:1985}
{Amari S.-i.}, 1985, \textit{Differential--geometrical methods in statistics},
  Lecture Notes in Statistics \textbf{28}, Springer, Berlin.

\bibitem{Amari:2009:alpha:divergence}
{Amari S.-i.}, 2009, \textit{$\alpha$-divergence is unique, belonging to both
  $f$-divergence and Bregman divergence classes}, IEEE Trans. Inf. Theor.
  \textbf{55}, 4925.

\bibitem{Amari:Nagaoka:1993}
{Amari S.-i., Nagaoka H.}, 1993, \textit{Joho kika no hoho}, Iwanami Shoten,
  T\={o}ky\={o} (engl. transl. rev. ed.: 2000, \textit{Methods of information
  geometry}, American Mathematical Society, Providence).

\bibitem{Araki:1976:relative:entropy:I}
{Araki H.}, 1976, \textit{Relative entropy for states of von Neumann algebras
  I}, Publ. Res. Inst. Math. Sci. Ky\={o}to Univ. \textbf{11}, 809. Available
  at:
  \href{http://www.ems-ph.org/journals/show_pdf.php?issn=0034-5318&vol=11&iss=3&rank=9}{www.ems-ph.org/journals/show\_pdf.php?issn=0034-5318\&vol=11\&iss=3\&rank=9}.

\bibitem{Araki:1977:relative:entropy:II}
{Araki H.}, 1977, \textit{Relative entropy for states of von Neumann algebras
  II}, Publ. Res. Inst. Math. Sci. Ky\={o}to Univ. \textbf{13}, 173. Available
  at:
  \href{http://www.ems-ph.org/journals/show_pdf.php?issn=0034-5318&vol=13&iss=1&rank=8}{www.ems-ph.org/journals/show\_pdf.php?issn=0034-5318\&vol=13\&iss=1\&rank=8}.

\bibitem{Araki:1999}
{Araki H.}, 1993, \textit{Ryoshiba no suri}, Iwanami Shoten, T\={o}ky\={o}
  (engl. transl. 1999, \textit{Mathematical theory of quantum fields}, Int.
  Ser. Monogr. Phys. \textbf{101}, Oxford University Press, Oxford).

\bibitem{BGW:2005}
{Banerjee A., Guo X., Wang H.}, 2005, \textit{On the optimality of conditional
  expectation as a Bregman predictor}, IEEE Trans. Inf. Theor. \textbf{51},
  2664.

\bibitem{Bauer:1958}
{Bauer H.}, 1958, \textit{Minimalstellen von Funktionen und Extremalpunkte},
  Arch. Math. \textbf{9}, 389.

\bibitem{Baumgaertel:1995}
{Baumg\"{a}rtel H.}, 1995, \textit{Operatoralgebraic methods in quantum field
  theory}, Akademie Verlag, Berlin.

\bibitem{Bauschke:Borwein:1997}
{Bauschke H.H., Borwein J.M.}, 1997, \textit{Legendre functions and the method
  of random Bregman projections}, J. Conv. Anal. \textbf{4}, 27. Available at:
  \href{http://www.emis.de/journals/JCA/vol.4_no.1/j86.ps.gz}{www.emis.de/journals/JCA/vol.4\_no.1/j86.ps.gz}.

\bibitem{Bayes:1763}
{Bayes T.}, 1763, \textit{An essay towards solving a problem in the doctrine of
  chances}, Phil. Trans. Roy. Soc. London \textbf{53}, 370 (reprinted in: 1958,
  Biometrika \textbf{45}, 293).

\bibitem{Borel:1909:denom}
{Borel \'{E}.}, 1909, \textit{Les probabilit\'{e}s d\'{e}nomerables et leurs
  applications arithm\'{e}tiques}, Rend. Circ. Math. Palermo \textbf{27}, 247.

\bibitem{Bratelli:Robinson:1979}
{Bratelli O., Robinson D.W.}, 1979, 1981, \textit{Operator algebras and quantum
  statistical mechanics}, Vol.1-2, Springer, Berlin.

\bibitem{Bregman:1967}
{Bregman L.M.}, 1967, \textit{Relaksatsionny\v{\i} metod nakhozdheniya
  obshche\v{\i} tochki vypuklykh mnozhestv i ego primenenie dla zadach
  vypuklogo programmirovaniya}, Zh. vychestl. matem. matem. fiz. \textbf{7},
  620. (engl. transl.: 1967, \textit{The relaxation method for finding common
  points of convex sets and its application to the solution of problems in
  convex programming}, USSR Comput. Math. Math. Phys. \textbf{7}, 200.).

\bibitem{Busch:Grabowski:Lahti:1989}
{Busch P., Grabowski M., Lahti P.J.}, 1989, \textit{Some remarks on effects,
  operations, and unsharp measurements}, Found. Phys. Lett. \textbf{2}, 331.

\bibitem{Busch:Grabowski:Lahti:1995}
{Busch P., Grabowski M., Lahti P.J.}, 1995, \textit{Operational quantum
  physics}, LNPm \textbf{31}, Springer, Berlin. (1997, 2nd corr. print.).

\bibitem{Caticha:1998:PRA}
{Caticha A.}, 1998, \textit{Consistency, amplitudes and probabilities in
  quantum theory}, Phys. Rev. A \textbf{57}, 1572.
  \href{http://www.arxiv.org/pdf/quant-ph/9804012}{arXiv:quant-ph/9804012}.

\bibitem{Caticha:1998:PLA}
{Caticha A.}, 1998, \textit{Consistency and linearity in quantum theory}, Phys.
  Lett. A \textbf{244}, 13.
  \href{http://www.arxiv.org/pdf/quant-ph/9803086}{arXiv:quant-ph/9803086}.

\bibitem{Caticha:2000:insufficient}
{Caticha A.}, 2000, \textit{Insufficient reason and entropy in quantum theory},
  Found. Phys. \textbf{30}, 227.
  \href{http://www.arxiv.org/pdf/quant-ph/9810074}{arXiv:quant-ph/9810074}.

\bibitem{Caticha:Giffin:2006}
{Caticha A., Giffin A.}, 2006, \textit{Updating probabilities}, in:
  Mohammad-Djafari A. (ed.), \textit{Bayesian inference and maximum entropy
  methods in science and engineering}, AIP Conf. Proc. \textbf{872},
  31.\href{http://www.arxiv.org/pdf/physics/0608185}{arXiv:physics/0608185}.

\bibitem{Caves:Fuchs:Schack:2001}
{Caves C.M., Fuchs C.A., Schack R.}, 2001, \textit{Quantum probabilities as
  bayesian probabilities}, Phys. Rev. A \textbf{65}, 022305.
  \href{http://www.arxiv.org/pdf/quant-ph/0106133}{arXiv:quant-ph/0106133}.

\bibitem{Chentsov:1968}
{Chentsov N.N.}, 1968, \textit{Nesimmetrichnoye rasstoyanie mezhdu
  raspredeleniyami veroyatnoste\v{\i}, entropiya i teoriya Pifagora}, Mat.
  Zametki \textbf{4}, 323. (engl. transl. 1968, \textit{Nonsymmetrical distance
  between probability distributions, entropy and the theorem of Pythagoras},
  Math. Notes \textbf{4}, 686.).

\bibitem{Chentsov:1972}
{Chentsov N.N.}, 1972, \textit{Statisticheskie reshayushchie pravila i
  optimal'nye vyvody}, Nauka, Moskva (engl. transl.: 1982, \textit{Statistical
  decision rules and optimal inference}, American Mathematical Society,
  Providence).

\bibitem{Connes:1973:classification}
{Connes A.}, 1973, \textit{Une classification des facteurs de type III}, Ann.
  Sci. \'{E}cole Norm. Sup. 4 \`{e}me s\'{e}r. \textbf{6}, 133.
  \href{http://archive.numdam.org/article/ASENS_1973_4_6_2_133_0.pdf}{numdam:ASENS\_1973\_4\_6\_2\_133\_0}.

\bibitem{Cox:1946}
{Cox R.T.}, 1946, \textit{Probability, frequency and reasonable expectation},
  Am. J. Phys. \textbf{14}, 1.

\bibitem{Cox:1961}
{Cox R.T.}, 1961, \textit{The algebra of probable inference}, John Hopkins
  University Press, Baltimore.

\bibitem{Csiszar:1963}
{Csisz\'{a}r I.}, 1963, \textit{Eine informationstheorische Ungleichung und
  ihre Anwendung auf den Beweis der Ergodizit\"{a}t von Markoffschen Ketten},
  Magyar Tud. Akad. Mat. Kutat\'{o} Int. K\"{o}zl. \textbf{8}, 85.

\bibitem{Csiszar:1975}
{Csisz\'{a}r I.}, 1975, \textit{$I$-divergence geometry of probability
  distributions and minimization problems}, Ann. Prob. \textbf{3}, 146.

\bibitem{Csiszar:1978}
{Csisz\'{a}r I.}, 1978, \textit{Information measures: a critical survey}, in:
  Kozesnik J. (ed.), \textit{Transactions of the seventh Prague conference on
  information theory, statistical decision functions and random processes},
  p.73.

\bibitem{Csiszar:1991}
{Csisz\'{a}r I.}, 1991, \textit{Why least squares and maximum entropy? An
  axiomatic approach to inference for linear inverse problems}, Ann. Stat.
  \textbf{19}, 2032.

\bibitem{Daniell:1918}
{Daniell P.J.}, 1918, \textit{A general form of the integral}, Ann. Math.
  \textbf{19}, 274.

\bibitem{Daniell:1919:a}
{Daniell P.J.}, 1919, \textit{Integrals in an infinite number of dimensions},
  Ann. Math. \textbf{21}, 30.

\bibitem{Daniell:1920}
{Daniell P.J.}, 1920, \textit{Further properties of the general integral}, Ann.
  Math \textbf{21}, 203.

\bibitem{deFinetti:1931}
{de Finetti B.}, 1931, \textit{Sul significato soggestivo della
  probabilit\`{a}}, Mathematicae \textbf{17}, 298.

\bibitem{deFinetti:1970}
{de Finetti B.}, 1970, \textit{Teoria delle probabilit\`{a}}, Einaudi, Torino
  (engl. transl.: 1974, \textit{Theory of probability}, Vol.1-2, Wiley, New
  York.

\bibitem{Laplace:1812}
{de Laplace P.-S.}, 1812, \textit{Th\'{e}orie analytique des probabilit\'{e}s},
  Courcier, Paris.

\bibitem{deMuynck:2002}
{de Muynck W.M.}, 2002, \textit{Foundations of quantum mechanics: an empiricist
  approach}, Kluwer, Dordrecht.

\bibitem{Eguchi:1983}
{Eguchi S.}, 1983, \textit{Second order efficiency of minimum contrast
  estimators in a curved exponential family}, Ann. Statist. \textbf{11}, 793.
  \href{http://projecteuclid.org/DPubS/Repository/1.0/Disseminate?view=body&id=pdf_1&handle=euclid.aos/1176346246}{euclid:1176346246}.

\bibitem{Eguchi:1985}
{Eguchi S.}, 1985, \textit{A differential geometric approach to statistical
  inference on the basis of contrast functionals}, Hiroshima Math. J.
  \textbf{15}, 341.
  \href{http://projecteuclid.org/DPubS/Repository/1.0/Disseminate?view=body&id=pdf_1&handle=euclid.hmj/1206130775}{euclid:1206130775}.

\bibitem{Emch:1972}
{Emch G.G.}, 1972, \textit{Algebraic methods in statistical mechanics and
  quantum field theory}, Wiley, New York.

\bibitem{Emch:1998}
{Emch G.G.}, 1998, \textit{On the need to adapt de Finetti's probability
  interpretation to QM}, Banach Center Publ. \textbf{43}, 157.

\bibitem{Falcone:Takesaki:2001}
{Falcone T., Takesaki M.}, 2001, \textit{The non-commutative flow of weights on
  a von Neumann algebra}, J. Funct. Anal. \textbf{182}, 170. Available at:
  \href{http://www.math.ucla.edu/~mt/papers/QFlow-Final.tex.pdf}{www.math.ucla.edu/$\sim$mt/papers/QFlow-Final.tex.pdf}.

\bibitem{Fremlin:2000}
{Fremlin D.H.}, 2000, 2001, 2002, 2003, \textit{Measure theory}, Vol.1-4,
  Torres Fremlin, Colchester.

\bibitem{Fuchs:2003}
{Fuchs C.A.}, 2003, \textit{Quantum mechanics as quantum information, mostly},
  J. Mod. Opt. \textbf{50}, 987.
  \href{http://www.arxiv.org/pdf/quant-ph/0205039}{arXiv:quant-ph/0205039}.

\bibitem{Fuchs:2010}
{Fuchs C.A.}, 2010, \textit{Qbism, the perimeter of quantum bayesianism},
  \href{http://www.arxiv.org/pdf/1003.5209}{arXiv:1003.5209}.

\bibitem{Gibilisco:Isola:1999}
{Gibilisco P., Isola T.}, 1999, \textit{Connections on statistical manifolds of
  density operators by geometry of non-commutative $L^{p}$ spaces}, Infin. Dim.
  Anal. Quant. Prob. Relat. Topics \textbf{2}, 169. Available at:
  \href{http://www.mat.uniroma2.it/~isola/research/preprints/GiIs01.pdf}{www.mat.uniroma2.it/$\sim$isola/research/preprints/GiIs01.pdf}.

\bibitem{Gibilisco:Pistone:1998}
{Gibilisco P., Pistone G.}, 1998, \textit{Connections on non-parametric
  statistical manifolds by Orlicz space geometry}, Inf. Dim. Anal. Quant. Prob.
  Relat. Top. \textbf{1}, 325.

\bibitem{Giffin:2009}
{Giffin A.}, 2009, \textit{Maximum entropy: the universal method for
  inference}, Ph.D. thesis, State University of New York, Albany.
  \href{http://www.arxiv.org/pdf/0901.2987}{arXiv:0901.2987}.

\bibitem{Gleason:1957}
{Gleason A.}, 1957, \textit{Measures on the closed subspaces of a Hilbert
  space}, J. Math. Mech. \textbf{6}, 885.

\bibitem{Grabowski:1989}
{Grabowski M.}, 1989, \textit{What is an observable?}, Fund. Phys. \textbf{19},
  923.

\bibitem{Grandy:1987}
{Grandy W.T., Jr.}, 1987, 1988, \textit{Foundations of statistical mechanics},
  Vol.1-2, Reidel, Dordrecht.

\bibitem{Grandy:2008}
{Grandy W.T., Jr.}, 2008, \textit{Entropy and the time evolution of macroscopic
  systems}, Oxford University Press, Oxford.

\bibitem{Haag:1959}
{Haag R.}, 1959, \textit{Discussion des ``axiomes'' et des propri\'{e}tes
  asymptotiques d'une th\'{e}orie des champs locale avec particules
  compos\'{e}es}, in: ? (ed.), \textit{Les probl\`{e}mes math\'{e}matique de la
  th\'{e}orie quantique des champs (Lille 1957)}, Centr. Natl. Rech. Sci.,
  Paris..

\bibitem{Haag:1992}
{Haag R.}, 1992, \textit{Local quantum physics}, Springer, Berlin. (1996, 2nd
  rev. enlarg. ed.).

\bibitem{Haagerup:1975:standard:form}
{Haagerup U.}, 1975, \textit{The standard form of von Neumann algebras}, Math.
  Scand. \textbf{37}, 271.

\bibitem{Hamhalter:2003}
{Hamhalter J.}, 2003, \textit{Quantum measure theory}, Kluwer, Dordrecht.

\bibitem{Hasegawa:1993}
{Hasegawa H.}, 1993, \textit{$\alpha$-divergence of the non-commutative
  information geometry}, Rep. Math. Phys. \textbf{33}, 87.

\bibitem{Ingarden:1963}
{Ingarden R.S.}, 1963, \textit{Information theory and variational principles in
  statistical theories}, Bull. Acad. Polon. Sci. S\'{e}r. Math. Astr. Phys.
  \textbf{11}, 541.

\bibitem{Ingarden:1976}
{Ingarden R.S.}, 1976, \textit{Quantum information theory}, Rep. Math. Phys.
  \textbf{10}, 43.

\bibitem{Ingarden:1978}
{Ingarden R.S.}, 1978, \textit{Information thermodynamics and differential
  geometry}, Memoirs Sagami Inst. Techn. \textbf{12}, 83.

\bibitem{Ingarden:Janyszek:1982}
{Ingarden R.S., Janyszek H.}, 1982, \textit{On the local riemannian structure
  of the state space of classical information thermodynamics}, Tensor (N.S.)
  \textbf{39}, 279.

\bibitem{IJKK:1982}
{Ingarden R.S., Janyszek H., Kossakowski A., Kawaguchi T.}, 1982,
  \textit{Information geometry of quantum statistical systems}, Tensor (N.S.)
  \textbf{37}, 105.

\bibitem{IKO:1997}
{Ingarden R.S., Kossakowski A., Ohya M.}, 1997, \textit{Information dynamics
  and open systems -- classical and quantum approach}, Kluwer, Dordrecht.

\bibitem{Ingarden:Urbanik:1961}
{Ingarden R.S., Urbanik K.}, 1961, \textit{Information as a fundamental notion
  of statistical physics}, Bull. Acad. Polon. Sci., S\'{e}r. Sci. Math.
  Astronom. Phys. \textbf{9}, 313.

\bibitem{Ingarden:Urbanik:1962}
{Ingarden R.S., Urbanik K.}, 1962, \textit{Information without probability},
  Colloq. Math. \textbf{9}, 131.

\bibitem{Jaynes:1957}
{Jaynes E.T.}, 1957, \textit{Information theory and statistical mechanics},
  Phys. Rev. \textbf{106}, 620.

\bibitem{Jaynes:1957:2}
{Jaynes E.T.}, 1957, \textit{Information theory and statistical mechanics II},
  Phys. Rev. \textbf{108}, 171.

\bibitem{Jaynes:1967}
{Jaynes E.T.}, 1967, \textit{Foundations of probability theory and statistical
  mechanics}, in: Bunge M. (ed.), \textit{Delaware seminars in the foundations
  of science}, Vol.1, Springer, Berlin, p.77.

\bibitem{Jaynes:1979:where:do:we:stand}
{Jaynes E.T.}, 1979, \textit{Where do we stand on maximum entropy?}, in: Levine
  R.D., Tribus M. (eds.), \textit{The maximum entropy formalism}, MIT Press,
  Cambridge, p.15.

\bibitem{Jaynes:1985:scattering}
{Jaynes E.T.}, 1985, \textit{Generalized scattering}, in: Smith C.R., Grandy
  W.T., Jr. (eds.), \textit{Maximum--entropy and bayesian methods in inverse
  problems}, Reidel, Dordrecht, p.377.

\bibitem{Jaynes:1985:macropred}
{Jaynes E.T.}, 1985, \textit{Macroscopic prediction}, in: Haken H. (ed.),
  \textit{Complex systems -- operational approaches in neurobiology, physics
  and computers}, Springer, Berlin, p.254.

\bibitem{Jaynes:1986}
{Jaynes E.T.}, 1986, \textit{Predictive statistical mechanics}, in: Moore G.T.,
  Scully M.O. (eds.), \textit{Frontiers of nonequilibrium statistical
  mechanics}, Plenum, New York, p.33.

\bibitem{Jaynes:1990}
{Jaynes E.T.}, 1990, \textit{Probability in quantum theory}, in: \.Zurek W.H.
  (ed.), \textit{Complexity, entropy and the physics of onformation}, Addison
  Wesley, Reading.

\bibitem{Jaynes:1993}
{Jaynes E.T.}, 1993, \textit{Inferential scattering}, unpublished preprint,
  available at
  \href{http://bayes.wustl.edu/etj/articles/cinfscat.pdf}{bayes.wustl.edu/etj/articles/cinfscat.pdf}.

\bibitem{Jaynes:2003}
{Jaynes E.T.}, 2003, \textit{Probability theory: the logic of science},
  Cambridge University Press, Cambridge.

\bibitem{Jaynes:Scalapino:1963}
{Jaynes E.T., Scalapino D.J.}, 1963, \textit{Irreversible statistical
  mechanics}, unpublished preprint, available at:
  \href{http://bayes.wustl.edu/etj/articles/irreversible.stat.mac.pdf}{bayes.wustl.edu/etj/articles/irreversible.stat.mac.pdf}.

\bibitem{Jencova:2003:arXiv}
{Jen\v{c}ov\'{a} A.}, 2003, \textit{Affine connections, duality and divergences
  for a von Neumann algebra},
  \href{http://arxiv.org/PS_cache/math-ph/pdf/0311/0311004v1.pdf}{arXiv:math-ph/0311004}.

\bibitem{Jencova:2005}
{Jen\v{c}ov\'{a} A.}, 2005, \textit{Quantum information geometry and
  non-commutative $L_p$ spaces}, Inf. Dim. Anal. Quant. Prob. Relat. Top.
  \textbf{8}, 215. Available at:
  \href{http://www.mat.savba.sk/~jencova/lpspaces.pdf}{www.mat.savba.sk/$\sim$jencova/lpspaces.pdf}.

\bibitem{Jencova:2006}
{Jen\v{c}ov\'{a} A.}, 2006, \textit{A construction of a nonparametric quantum
  information manifold}, J. Funct. Anal. \textbf{239}, 1.
  \href{http://www.arxiv.org/pdf/math-ph/0511065}{arXiv:math-ph/0511065}.

\bibitem{Jencova:2010}
{Jen\v{c}ov\'{a} A.}, 2010, \textit{On quantum information manifolds}, in:
  Gibilisco P. et al (eds.), \textit{Algebraic and geometric methods in
  statistics}, Cambridge University Press, Cambridge,.

\bibitem{Kalashnikov:Zubarev:1972}
{Kalashnikov V.P., Zubarev D.N.}, 1972, \textit{On the extremal properties of
  the nonequilibrium statistical operator}, Physica \textbf{59}, 314.

\bibitem{Klauder:1997}
{Klauder J.R.}, 1997, \textit{Coherent states in action}, in: Lim S.C. et al
  (eds.), \textit{Frontiers in quantum physics}, Springer, Berlin, p.151.
  \href{http://www.arxiv.org/pdf/quant-ph/9710029}{arXiv:quant-ph/9710029}.

\bibitem{Kolmogorov:1933}
{Kolmogorov A.N.}, 1933, \textit{Grundbergriffe der
  Wahrscheinlichkeitrechnung}, Ergebnisse der Mathematik und Ihrer Grenzgebiete
  Bd. \textbf{2}, Springer, Berlin. (engl. transl. 1950, \textit{Foundations of
  the theory of probability}, Chelsea, New York).

\bibitem{Kostecki:2011:OSID}
{Kostecki R.P.}, 2011, \textit{The general form of $\gamma$-family of quantum
  relative entropies}, Open Sys. Inf. Dyn. \textbf{18}, 191.
  \href{http://www.arxiv.org/pdf/1106.2225}{arXiv:1106.2225}.

\bibitem{Kostecki:2011:QIG}
{Kostecki R.P.}, 2011, \textit{On general form of quantum information
  geometry}, to be submitted.

\bibitem{Kostecki:2011:principles}
{Kostecki R.P.}, 2011, \textit{On principles of inductive inference}, submitted
  to: \textit{Proceedings of 31st International Workshop on Bayesian Inference
  and Maximum Entropy Methods in Science and Engineering, 10-15 July 2011,
  Waterloo}.

\bibitem{Kostecki:2011:towards:series}
{Kostecki R.P.}, 2011, \textit{Towards new foundations of quantum theory I, II,
  III}, in preparation.

\bibitem{LeCam:1964}
{Le Cam L.}, 1964, \textit{Sufficiency and approximate sufficiency}, Ann. Math.
  Statist. \textbf{35}, 1419.

\bibitem{LeCam:1986}
{Le Cam L.}, 1986, \textit{Asymptotic methods in statistical decision theory},
  Springer, Berlin.

\bibitem{Loomis:1947}
{Loomis L.H.}, 1947, \textit{On the representation of $\sigma$-complete boolean
  algebras}, Bull. Amer. Math. Soc. \textbf{53}, 757.

\bibitem{Maeda:1990}
{Maeda S.}, 1990, \textit{Probability measures on projections in von Neumann
  algebras}, Rev. Math. Phys. \textbf{1}, 235..

\bibitem{Mitchell:1967}
{Mitchell W.C.}, 1967, \textit{Statistical mechanics of thermally driven
  systems}, Ph.D. thesis, Washington University, St.Louis.

\bibitem{Morimoto:1963}
{Morimoto T.}, 1963, \textit{Markov processes and the $H$-theorem}, J. Phys.
  Soc. Jap. \textbf{12}, 328.

\bibitem{Morozov:Roepke:1998}
{Morozov V.G., R\"{o}pke G.}, 1998, \textit{Zubarev's method of a
  nonequilibrium statistical operator and some challenges in the theory of
  irreversible processes}, Cond. Matt. Phys. \textbf{16}, 673.

\bibitem{Morozova:Chentsov:1991}
{Morozova E.A., Chentsov N.N.}, 1991, \textit{Estestvennaya geometriya
  seme\v{\i}stv veroyatnostnykh zakonov}, Itogi Nauki i Tekhniki, Ser. Sovr.
  Probl. Mat. Fundam. Napravlen. \textbf{83}, 133. (\textit{The natural
  geometry of families of probabilistic laws}) Available at:
  \href{http://www.mathnet.ru/links/0b7482826af98f74f707088ff33082f6/intf210.pdf}{www.mathnet.ru/links/0b7482826af98f74f707088ff33082f6/intf210.pdf}.

\bibitem{Nagaoka:Amari:1982}
{Nagaoka H., Amari S.-i.}, 1982, \textit{Differential geometry of smooth
  families of probability distributions}, Technical report METR \textbf{82-7},
  University of T\={o}ky\={o}, T\={o}ky\={o}.

\bibitem{Neumaier:2003}
{Neumaier A.}, 2003, \textit{Ensembles and experiments in classical and quantum
  physics}, Int. J. Mod. Phys. B \textbf{17}, 2937.
  \href{http://www.arxiv.org/pdf/quant-ph/0303047}{arXiv:quant-ph/0303047}.

\bibitem{Ohya:Petz:1993}
{Ohya M., Petz D.}, 1993, \textit{Quantum entropy and its use}, Springer,
  Berlin.

\bibitem{Ojima:2004}
{Ojima I.}, 2004, \textit{Temperature as order parameter of broken scale
  invariance}, Publ. Res. Inst. Math. Sci. Ky\={o}to Univ. \textbf{40}, 731.
  \href{http://arxiv.org/PS_cache/math-ph/pdf/0311/0311025v1.pdf}{arXiv:math-ph/0311025}.

\bibitem{Petz:1985:quasientropies}
{Petz D.}, 1985, \textit{Quasi-entropies for states of a von Neumann algebra},
  Publ. Res. Inst. Math. Sci. Ky\={o}to Univ. \textbf{21}, 787. Available at:
  \href{http://www.math.bme.hu/~petz/pdf/26quasi.pdf}{www.math.bme.hu/$\sim$petz/pdf/26quasi.pdf}.

\bibitem{Petz:1986:properties}
{Petz D.}, 1986, \textit{Properties of the relative entropy of states of von
  Neumann algebra}, Acta Math. Hungar. \textbf{47}, 65.

\bibitem{Petz:2007:Bregman}
{Petz D.}, 2007, \textit{Bregman divergence as relative operator entropy}, Acta
  Math. Hungar. \textbf{116}, 127. Available at:
  \href{http://www.renyi.hu/~petz/pdf/112bregman.pdf}{www.renyi.hu/$\sim$petz/pdf/112bregman.pdf}.

\bibitem{Piron:1976}
{Piron C.}, 1976, \textit{Foundations of quantum physics}, Benjamin, New York.

\bibitem{Pistone:Sempi:1995}
{Pistone G., Sempi C.}, 1995, \textit{An infinite-dimensional geometric
  structure on the space of all the probability measures equivalent to a given
  one}, Ann. Statist. \textbf{23}, 1543.

\bibitem{Pitowsky:2006}
{Pitowsky I.}, 2006, \textit{Quantum mechanics as a theory of probability}, in:
  Demopoulos W., Pitowsky I. (eds.), \textit{Physical theory and its
  interpretation: essays in honor of Jeffrey Bub}, Springer, Berlin, p.213.
  \href{http://arxiv.org/PS_cache/quant-ph/pdf/0510/0510095v1.pdf}{arXiv:quant-ph/0510095}.

\bibitem{Ramsey:1931}
{Ramsey F.P.}, 1931, \textit{The foundations of mathematics and other logical
  essays}, Routledge and Kegan, London.

\bibitem{Randall:Foulis:1983}
{Randall C.H., Foulis D.J.}, 1983, \textit{Properties and operational
  propositions in quantum mechanics}, Found. Phys. \textbf{13}, 843.

\bibitem{Redei:Summers:2007}
{R\'{e}dei M., Summers S.J.}, 2007, \textit{Quantum probability theory}, Stud.
  Hist. Phil. Mod. Phys. \textbf{38}, 390.
  \href{http://www.arxiv.org/pdf/quant-ph/0601158}{arXiv:quant-ph/0601158}.

\bibitem{Segal:1947:postulates}
{Segal I.E.}, 1947, \textit{Postulates for general quantum mechanics}, Ann.
  Math. \textbf{48}, 930.

\bibitem{Segal:1953}
{Segal I.E.}, 1953, \textit{A non-commutative extension of abstract
  integration}, Ann. Math. \textbf{57}, 401.

\bibitem{Shannon:1948}
{Shannon C.}, 1948, \textit{A mathematical theory of communication}, Bell Syst.
  Tech. J. \textbf{27}, 379, 623.

\bibitem{Sikorski:1948}
{Sikorski R.}, 1948, \textit{On the representation of boolean algebras as
  fields of sets}, Fund. Math. \textbf{35}, 247.

\bibitem{Stone:1948}
{Stone M.H.}, 1948, \textit{Notes on integration I, II, III}, Proc. Natl. Acad.
  Sci. U.S.A. \textbf{34}, 336, 447, 483.

\bibitem{Stone:1949}
{Stone M.H.}, 1949, \textit{Notes on integration IV}, Proc. Natl. Acad. Sci.
  U.S.A. \textbf{35}, 50.

\bibitem{Umegaki:1962}
{Umegaki H.}, 1962, \textit{Conditional expectations in an operator algebra IV
  (entropy and information)}, K\={o}dai Math. Sem. Rep. \textbf{14}, 59.

\bibitem{Whittle:1970}
{Whittle P.}, 1970, \textit{Probability}, Penguin, Harmondsworth.

\bibitem{Whittle:2000}
{Whittle P.}, 2000, \textit{Probability via expectation}, (4th ed.), Springer,
  Berlin.

\bibitem{Wigner:Yanase:1963}
{Wigner E., Yanase M.}, 1963, \textit{Information content of distributions},
  Proc. Nat. Acad. Sci. U.S.A. \textbf{49}, 910.

\bibitem{Wilce:2000}
{Wilce A.}, 2000, \textit{Test spaces and orthoalgebras}, in: Coecke B., Moore
  D., Wilce A. (eds.), \textit{Current research in operational quantum logic},
  Kluwer, Dordrecht, p.?.

\bibitem{Youssef:1991}
{Youssef S.}, 1991, \textit{A reformulation of quantum mechanics}, Mod. Phys.
  Lett. A \textbf{6}, 225.

\bibitem{Youssef:1994}
{Youssef S.}, 1994, \textit{Quantum mechanics as complex probability theory},
  Mod. Phys. Lett. A \textbf{9}, 2571.

\bibitem{Zhang:2004:divergence}
{Zhang J.}, 2004, \textit{Divergence function, duality, and convex analysis},
  Neural Comp. \textbf{16}, 159. Available at:
  \href{http://www.lsa.umich.edu/psych/junz/Neural%20Comp%202004.pdf}{www.lsa.umich.edu/psych/junz/Neural\%20Comp\%202004.pdf}.

\bibitem{Zhu:1998:lebesgue}
{Zhu H.}, 1998, \textit{Generalized Lebesgue spaces and application to
  statistics}, Santa Fe Inst. Tech. Rep. \textbf{98-06-44}, Santa Fe. Available
  at:
  \href{http://www.santafe.edu/media/workingpapers/98-06-044.ps}{www.santafe.edu/media/workingpapers/98-06-044.ps}.

\bibitem{Zhu:Rohwer:1997}
{Zhu H., Rohwer R.}, 1997, \textit{Measurements of generalisation based on
  information geometry}, in: Ellacott S.W. et al. (eds.), \textit{Mathematics
  of neural networks: models, algorithms and applications}, Kluwer, Dordrecht,
  p.394. Available at:
  \href{http://eprints.aston.ac.uk/514/1/NCRG_95_012.pdf}{eprints.aston.ac.uk/514/1/NCRG\_95\_012.pdf}.

\bibitem{Zhu:Rohwer:1998}
{Zhu H., Rohwer R.}, 1998, \textit{Information geometry, bayesian inference,
  ideal estimates and error decomposition}, Santa Fe Inst. Tech. Rep.
  \textbf{98-06-45}, Santa Fe. Available at:
  \href{http://omega.albany.edu:8008/ignorance/zhu98.pdf}{omega.albany.edu:8008/ignorance/zhu98.pdf}.

\bibitem{Zubarev:1961}
{Zubarev D.N.}, 1961, \textit{[Statisticheski\v{\i} operator dla neravnovesnykh
  sistem]}, Dokl. Akad. Nauk SSSR \textbf{140}, 92. (engl. transl. 1962,
  \textit{The statistical operator for nonequilibrium systems}, Sov. Phys.
  Dokl. \textbf{6}, 776).

\bibitem{Zubarev:1994}
{Zubarev D.N.}, 1994, \textit{Nonequilibrium statistical operator as a
  generalization of Gibbs distribution for nonequilibrium case}, Cond. Matt.
  Phys. \textbf{4}, 7.

\bibitem{Zubarev:Kalashnikov:1969}
{Zubarev D.N., Kalashnikov V.P.}, 1969, \textit{Ekstremalnye svo\v{\i}stva
  neravnovestnogo statisticheskogo operatora}, Teor. Mat. Fiz. \textbf{1}, 137.
  (\textit{Extremal properties of the nonequilibrium statistical operator}).

\bibitem{Zubarev:Kalashnikov:1970:Physica}
{Zubarev D.N., Kalashnikov V.P.}, 1970, \textit{Derivation of the
  nonequilibrium statistical operator from the extremum of the information
  entropy}, Physica \textbf{46}, 550.

\bibitem{Zubarev:Kalashnikov:1970:TMF}
{Zubarev D.N., Kalashnikov V.P.}, 1970, \textit{Postroyenne statisticheskikh
  operatorov dla neravnovesnykh processov}, Teor. Mat. Fiz. \textbf{3}, 126.
  (\textit{Construction of statistical operators for nonequilibrium
  processes}).

\bibitem{Zubarev:Morozov:Roepke:1996}
{Zubarev D.N., Morozov V.G., R\"{o}pke G.}, 1996, \textit{Statistical mechanics
  of nonequilibrium processes}, Vol.1-2, Akademie Verlag, Berlin. (2nd rev.
  enlarg. ed., 2002, \textit{Statisticheskaya mekhanika neravnovesnykh
  processov}, Vol.1-2, Fizmatlit, Moskva.).

\end{thebibliography}
}
\end{document}